# Seismological Research Letters

## The RESOLVE project: a multi-physics experiment with a temporary dense seismic array on the Argentière Glacier, French Alps

--Manuscript Draft--

|  | Fabian Walter |
| --- | --- |
| Order of Authors Secondary Information: |  |
| Suggested Reviewers: | Lukas Zoet<br>lzoet@wisc.edu |
|  | Marianne Karplus<br>mkarplus@utep.edu |
|  | Richard Aster<br>Rick.Aster@colostate.edu |
|  | Victor Tsai<br>victor_tsai@brown.edu |
|  | Paul Johnson<br>paj@lanl.gov |
|  |  |

|  | Nicolai Shapiro<br>nikolai.shapiro@univ-grenoble-alpes.fr |
| --- | --- |
|  | Adam Booth<br>A.D.Booth@leeds.ac.uk |
|  | Emma Smith<br>Emma.Smith@awi.de |

Letter to Editor   Click here to access/download;Letter to Editor;CoverLetter.pdf

Dear Editor,

Please find enclosed our manuscript entitled "The RESOLVE project: a multi-physics experiment with a temporary dense seismic array on the Argentière Glacier, French Alps" (Gimbert, F. et al.) submitted to the Data Mine columns of Seismological Research Letter.

Over the past decades seismic observations have yield important insights on glacier structure and dynamical processes (Podolskiy and Walter, 2016 Rev. Geo). However certain limitations have arisen as a result of the limited capability to retrieve the wavefield with sufficient spatial resolution and coverage to infer more quantitative information. Our manuscript documents a particularly novel experiment that aims to overcome these limitations. Our experiment consists in the deployment of a temporary dense seismic array on an Alpine Glacier. Such type of seismic monitoring is to our knowledge particularly unique given (i) the high number of deployed sensors (98), which enables monitoring with unprecedented density and coverage in such set-up, (ii) the acquiring of supplementary data providing key complementary constraints on glacier structure and dynamics, and (iii) the targeted site and time period of the year, where and during which key glacier structural and dynamical changes occur both in space and time and may be studied specifically with our array.

In the paper we first present our motivations for conducting such an experiment and we document the site and all aspects of our experiment. We then conduct preliminary analysis using multiple seismic techniques such as template matching and match field processing to illustrate the dataset and its capability to assess novel components of key glacier processes such as subglacial hydrology, stick slip motion and englacial fracturing. All data related to our experiment will be made available when the paper will be published, with all the necessary data access information being given in the *Data and Resources* section. We finally discuss in which ways further work using this dataset could to our view help tackle key remaining questions in the field.

We believe that our present study is a good illustration of an interdisciplinary work gathering scientists with different but complementary expertise targeting fundamental problems to address remaining challenges in cryo-seismology. Given the uniqueness of such type of experiment, we envision that it will be of particular interest to both geophysicists and glaciologists.

Thank you for your consideration of this paper.

Yours sincerely,

Florent Gimbert, on behalf of all co-authors.

Manuscript   Click here to access/download;Manuscript;RESOLVE_Draft_Submitted.pdf

# The RESOLVE project: a multi-physics experiment with a temporary dense seismic array on the Argentière Glacier, French Alps

Gimbert, F.[1], U. Nanni[1], P. Roux[2], A. Helmstetter[2], S. Garambois[2], A. Lecointre[2], A. 5 Walpersdorf[2], B. Jourdain[1], M. Langlais[2], O. Laarman[1], F. Lindner[3], A. Sergeant[4], C. Vincent[1], F. 6 Walter[3]

[1] University Grenoble Alpes, CNRS, IRD, IGE, Grenoble, France
[2] University Grenoble Alpes, CNRS, IRD, UGE, ISTerre, Grenoble, France
[3] Laboratory of Hydraulics, Hydrology and Glaciology, ETH Zürich, Zürich, Switzerland
[4] Aix Marseille Univ, CNRS, Centrale Marseille, LMA, France

**ABSTRACT**

Recent work in the field of cryo-seismology demonstrates that high frequency (>1 Hz) waves provide key constraints on a wide range of glacier processes such as basal friction, surface crevassing or subglacial water flow. Establishing quantitative links between the seismic signal and the processes of interest however requires detailed characterization of the wavefield, which at the high frequencies of interest necessitates the deployment of large and particularly dense seismic arrays. Although dense seismic array monitoring has recently become routine in geophysics, its application to glaciated environments has yet remained limited. Here we present a dense seismic array experiment made of 98 3-component seismic stations 23 continuously recording during 35 days in early spring on the Argentière Glacier, French Alps. The seismic dataset is supplemented by a wide range of complementary observations obtained from ground penetrating radar, drone imagery, GNSS positioning and in-situ instrumentation of basal glacier sliding velocities and subglacial water discharge. Through applying multiple processing techniques including event detection from template matching and systematic matched-field processing, we demonstrate that the present dataset provides enhanced spatial resolution on basal stick slip and englacial fracturing sources as well as novel constraints on the heterogeneous nature of the noise field generated by subglacial water flow and on the link between crevasse properties and englacial seismic velocities. We finally outline in which ways further work using this dataset could help tackle key remaining questions in the field.

**INTRODUCTION**

The deployment of large and dense seismic arrays has recently become routine in various geophysical contexts thanks to new technological developments of autonomous wireless geophones and increases in computational power. Spatially dense arrays allow enhancing the characterization of high frequency (>1 Hz) body waves and surface waves propagating in the subsurface, such as for example in near-surface fault systems exhibiting hundreds to few tens of meters long structures (e.g. the Newport-Inglewood Fault, see Lin *et al.* (2013), and the San Jacinto Fault, see Roux *et al.* (2016)). The improved resolution provided by dense arrays helps increase the completeness of impulsive seismic event catalogs (Vandemeulebrouck *et al.*, 2013), such that the spatio-temporal dynamics of sources may be studied in greater detail and event catalogs be used to conduct subsurface imaging (Meng and Ben-Zion, 2018; Chmiel *et al.*, 2019). Dense arrays also help to detect other sources of radiation (e.g. tremor and anthropogenic sources) compared to what is possible with single stations or regional networks (Inbal *et al.*, 2016; Li *et al.*, 2018; Meng and Ben-Zion, 2018).

Dense array monitoring techniques have however little been applied to the study of glaciers, although a number of seismic investigations have demonstrated that high frequency (> 1Hz) waves can provide key constraints on glacier dynamical processes and structure characteristics (Podolskiy and Walter, 2016). Analysis of impulsive sources yields insights on basal motion (Weaver and Malone, 1979; Allstadt and Malone, 2014; Helmstetter, Nicolas, *et al.*, 2015; Lipovsky and Dunham, 2016; Lipovsky *et al.*, 2019) and englacial fracturing (Neave and Savage, 1970; Roux *et al.*, 2010; Mikesell *et al.*, 2012; Podolskiy *et al.*, 2018; Garcia *et al.*, 2019). Continuous monitoring of englacial seismic velocities using background noise or impulsive arrivals yields insights on glacier structure such as the geometrical properties of surface or basal crevasses (Walter *et al.*, 2015; Lindner *et al.*, 2019; Zhan, 2019; Sergeant *et al.*, 2020). Analysis of continuous sources helps unravel the physics of subglacial water flow (Bartholomaus *et al.*, 2015; Gimbert *et al.*, 2016; Eibl *et al.*, 2020; Lindner *et al.*, 2020; Nanni *et al.*, 2020), the geometry and location of glacial moulins (Helmstetter, Moreau, *et al.*, 2015; Roeoesli *et al.*, 2016; Aso *et al.*, 2017) and the occurrence of subglacial sediment transport (Gimbert *et al.*, 2016).

The restricted use of dense array monitoring techniques in the above-mentioned applications may however limit our understanding on several aspects. The mechanisms responsible for stick-slip events involved in basal motion remain poorly identified, and the use of dense array monitoring techniques to more accurately infer where stick slip occurs and whether and at which rate stick-slip asperities migrate could help better infer the involved mechanisms and the necessary conditions for stick slip to operate (Lipovsky *et al.*, 2019). The relationship between crevasse characteristics such as depth or deformation rate and englacial seismic velocities is yet not fully established (Lindner *et al.*, 2019), and the use of dense arrays to concomitantly locate crevasse fracturing events and monitor changes in englacial seismic velocities along various directions could help establishing such relationship. The feasibility to infer the geometry of the subglacial drainage system as well as the dependency of subglacial water-flow-induced noise amplitudes and thus of inversions of subglacial flow physics (Gimbert *et al.*, 2016; Lindner *et al.*, 2020; Nanni *et al.*, 2020) on sensor position remain unexplored and could be addressed using dense array seismic observations.

Properly evaluating the knowledge gain dense seismic arrays may provide to address the above-mentioned challenges requires (i) monitoring a glacier that gathers all processes of interest, (ii) deploying an instrumentation that covers scales and durations over which significant changes operate, and (iii) acquiring complementary observations to test the seismically-derived findings and incorporate these into a wider glaciological context. Here we present data and preliminary analysis from a 98-sensors array deployed over 35 days during early spring 2018 on an Alpine Glacier, the Argentière Glacier in the French Alps (Fig. 1). We also provide and analyze key complementary observations from ground penetrating radar (GPR), drone imagery, Global National Satellite System (GNSS) positioning and in-situ instrumentation of basal glacier sliding velocities and subglacial water discharge. We argue that the selected glacier, the time period of investigation as well as the completeness of the present dataset allow satisfying all three above-mentioned conditions. Through applying multiple processing techniques including event detection from template matching and systematic matched-field processing, we demonstrate that the present dataset allows enhancing the spatial resolution associated with basal stick slip and near surface crevassing event locations as well as provides novel constraints on the degree of heterogeneity of the noise field caused by subglacial water flow and the variations of englacial seismic velocities.

We finally outline in which ways further work using this dataset could help overcome classical observational limitations and address key challenges in the field. The paper is structured as follows: we first describe the experimental strategy and setup, we then present preliminary seismic array results and we finally outline future studies that may be conducted in light of our preliminary results with this dataset.

**EXPERIMENT DESIGN**

FIELD SITE

The Argentière Glacier is located in the Mont Blanc Massif (French Alps, 45°55' N, 6°57'E, Fig. 1a) and is the second largest French glacier. It is about 10 km long, covers an area of about 12 km$^2$, and extends from an altitude of 1700 m asl up to about 3600 m asl. The upper part of the glacier is constricted in a typical U-shaped narrow valley where ice sits on granite. The lower part of the glacier is rather characterized by a sharper incised, V-shaped valley where ice sits on metamorphic rocks (Vallon, 1967; Hantz and Lliboutry, 1983; Vincent *et al.*, 2009). The glacier generally exhibits temperate bed conditions (Vivian and Bocquet, 1973), i.e. basal ice temperature is at the pressure melting point and water flow occurs at the interface as a result of being produced by year-round basal melt and summer surface melt (Cuffey and Paterson, 2010).

The monitored site is located in the lower part of the glacier (about 2 km from the glacier front) and at about 2400 m asl (Fig. 1a). In this area the surface slope is gentle (1-2%) and crevasses are restricted to an area of about 200 m from the glacier sides (Fig. 1b). The glacier flows at a year average velocity of about 60 m.yr$^{-1}$ in its center, about half of which is due to sliding at the ice-bed interface and the other half to internal ice deformation (Vincent and Moreau, 2016). Internal ice deformation in the area mainly occurs through ice creep, except near glacier sides where englacial fracturing may play a role. A strong seasonality is observed in glacier dynamics, with summer (typically May to September) velocities being equal to about 1.5 times winter velocities (Vincent and Moreau, 2016) as a result of melt water input lubricating the ice-bed interface and enhancing basal sliding (Lliboutry, 1959, 1968; Cuffey and Paterson, 2010).

The above-presented features of the Argentière Glacier make it an ideal case study to address our objectives of unravelling glacier structure and flow processes from seismic observations. Seismic studies over the past decade on this glacier have demonstrated that seismic observations are sensitive to surface crevassing (Helmstetter, Moreau, *et al.*, 2015), subglacial water flow (Nanni *et al.*, 2020), basal stick-slip (Helmstetter, Nicolas, *et al.*, 2015) and serac instabilities (Roux *et al.*, 2008). These processes are expected to generate a large panel of signals with broad azimuthal distributions and frequency contents that may be used for tomography.

SEISMIC INSTRUMENTATION AND GEOPHYSICAL CHARATERISATION OF GLACIER STRUCTURE AND DYNAMICS

**Seismic instrumentation**

Sensors of the dense seismic array (see red dots in Fig. 1b) are Fairfield ZLand 3 components nodes set up with a sampling frequency of 500 Hz (hereafter referred to as nodes). These sensors have a cut-off frequency of 4.5Hz, a sensitivity of 76.7 $V.m^{-1}.s^{-1}$ and a typical power autonomy of about 35 days. We deployed the nodes on April 24 when the glacier was entirely covered by a snow layer of about 3 m thick. We placed the sensors about 40 m apart from each other in the along-flow direction and about 50 m apart in the across-flow direction in order to enable subwavelength analysis in the 4-50 Hz frequency range of interest. We buried them into snow about 30 cm below the surface to ensure that sensors were levelled and well coupled over several days to a few weeks until snow melt uncovered it. This depth is also shallow enough to allow GNSS signal reception for time synchronization. Given that snow melt occurred at an average rate of about 2-3 $cm.day^{-1}$ during the investigated period and at this location, the 30-cm deep deployment necessitated sensors re-deployment once over the instrumented period, an operation that we conducted on May 11.

We supplemented the seismic array by one three-component borehole seismic station placed at 5 m below the ice-surface (see orange dot in Fig. 1c). This Geobit-C100 sensor connected to a Geobit-SRi32L digitizer provides higher sensitivity (1500 $V.m^{-1}.s^{-1}$), higher frequency sampling (1000 Hz) and a lower cut-off frequency (0.1 Hz) compared to the nodes. This seismic station is the same as the one used for the two-year long seismic study of Nanni *et al.* (2020).

**Recovery of surface and bed digital elevation models from structure from motion surveys and ground penetrating radar**

We construct a digital surface elevation model based on a drone geodetic survey that we conducted on September 5, 2018 when the glacier surface was snow free and crevasses could be identified. We used a senseFly eBee+ Unmanned Aerial Vehicle and acquired a total of 720 photos using the onboard senseFly S.O.D.A. camera (20 Mpx RGB sensor with 28 mm focal lens). We generate a digital elevation model of 10-cm resolution using differential Global Positioning System (GPS) measured ground control points (see green stars in Fig. 2a) and the Structure for Motion algorithm implemented in the software package Agisoft Metashape Professional version 1.5.2. A detailed description of the processing steps can be found in Brun *et al.* (2016) and Kraaijenbrink *et al.* (2016).

We calculate a crevasse map (see black dots in Fig. 2a) based on the surface digital elevation model, which has been shown to be more reliable and precise than using optical/radar images (Foroutan *et al.*, 2019). We first apply a 2D highpass filter with a low cut-off wavelength of 10 m and then define any location with elevation lower than -50 cm as being part of crevasses. Finally, we apply a 2D median filter with a 1 by 1 m kernel in order to remove artifacts from boulders and moraines.

To establish a digital elevation model of the glacier bed we primarily use Ground Penetrating Radar (GPR) data acquired using a system of two transmitting and receiving 4.2 MHz antennas connected to a time triggered acquisition developed especially for glacial applications by the Canadian company Blue System Integration Ltd (Mingo and Flowers, 2010). The GPR signal processing consists of correcting for source time excitation. We use both dynamic corrections to reproduce a zero-incidence acquisition from data acquired with a 20 m offset between source and receiver (Normal Moveout correction) and static corrections to highlight elevation variations along a profile. We do so using a constant wave velocity of $0.168 \cdot 10^9$ m.s$^{-1}$ that is typical for ice (Garambois *et al.*, 2016). We then apply a [1-15 MHz] Butterworth band-pass filter followed by a squared time gain amplification to the signal in order to increase signal-tonoise ratio. We show an illustration of the processed GPR data in Fig. 2b, where the direct airwave first arrival is followed by a large reflectivity V-shape pattern reaching $3000 \cdot 10^6$ s around the center of the profile. This latter profile corresponds to the ice/bedrock interface, although its apparent shape is biased by waves being reflected by the closest ice-bed interface rather than that located straight below the instrument. We correct for this bias by applying a frequency-wavenumber Stolt migration technique (Stolt, 1978) and converting time into distance using the constant wave velocity of $0.168 \cdot 10^6$ m.s$^{-1}$. We note that prior to migration we add null traces (i.e. with null amplitudes) in places where harsh glacier surface conditions (mainly crevasses) prevented us to acquire data. As illustrated in Fig. 2c the migration process is effective in correcting the artefacts due to the geometrical variation of the interface along the profile, which now appears as smooth and continuous. We then pick the ice-bed reflection (see yellow line in Fig. 2b) over all GPR profiles, such that a threedimensional bed DEM can be reconstructed.

We reconstruct a three-dimensional bed DEM over a larger area than that covered by GPR surveys by (i) incorporating additional constraints like glacier edge elevation as measured from drone imagery (see purple area in Fig. 2a) and in-situ borehole measured ice-bed interface elevations as obtained from the excavated tunnels located further down-glacier from the dense seismic array (see blue area in Fig. 2a) and (ii) interpolating all data using a kriging method onto on a 10*10 m grid. We estimate from different first onset pickings that the recovered depth uncertainty is of about 5 m below the seismic array, while we note that it likely is considerably larger and more on the order of few tens of meter outside of the array where observations are sparser.

In Fig. 2a we show the two-dimensional map of ice thickness as reconstructed based on subtracting the bed DEM from the surface DEM (using 25-m spaced contour lines). The glacier bed generally exhibits a gently dipping valley, with a maximum ice thickness of about 255 m at the center of the seismic array. Glacier thickness decreases relatively sharply on the glacier margins where surface crevasses are observed. We also observe that bed elevation significantly increases down glacier, which results in a decrease by more than 150 m in glacier thickness. Beyond these generic characteristics we identify two interesting reflectivity features in the migrated GPR images (see blue ellipses in Fig. 2c) that correspond to localized scattering observed near the surface and a large reflectivity pattern observed just above the deepest portion of the interface. The near surface scattering feature could be caused by deep crevasses, and the deeper feature could be caused by englacial and/or subglacial water conduits as recently proposed by Church *et al.* (2019), who made similar GPR observations in a temperate glacier and were able to verify such an interpretation from in-situ borehole observations.

**Meteorological and water discharge characteristics**

We use air temperature and precipitation measurements obtained at a 0.5 h time step with the automatic weather station maintained by the French glacier-monitoring program GLACIOCLIM (Les GLACIers un Observatoire du CLIMat; https://glacioclim.osug.fr/), which is located on the moraine next to the glacier at 2400 m asl (see green diamond in Fig. 1b). Precipitation is measured with an OTTPluvio weighing rain gauge. Water discharge routing subglacially is monitored at a 15 min time step in tunnels excavated into bedrock by the Emossons hydraulic power company, which are located 600 m downstream of the array center (at 2173 m asl) near the glacier ice fall (see blue star in Fig. 1b).

We can see that temperature generally increases over the instrumented period, from a multidaily average of about 0° C at the beginning of the measurement period to about 5 °C at the end (Fig. 3a). This general increase drives the general increase in water discharge, which varies from few tenths of $m^3.s^{-1}$ to several $m^3.s^{-1}$ over the period. Episodic rain events also occur during the instrumented period, but have little to no effect on subglacial discharge likely as a result of the snow cover acting as a buffer.

**Glacier dynamics instrumentation and general features**

We evaluate changes in glacier dynamics over the instrumented period by means of two observational methods. The first one is particularly unique to the present site, and consists of glacier basal sliding velocity measurements made continuously in the down glacier serac fall area (see red star in Fig. 1b) by means of a bicycle wheel placed directly in contact with the basal ice at the extremity of an excavated tunnel (Vivian and Bocquet, 1973; Vincent and Moreau, 2016). The wheel is coupled with a potentiometer that retrieves its rotation rate, which is then recorded digitally and converted back to a sliding velocity at a 1-s sampling time. The second type of measurements consists of 4 glacier surface and 1 reference bedrock GNSS stations (see yellow stars in Fig. 1b) of type Leica GR25 acquiring the GNSS signals every second. This temporary array is supplemented by a permanent ARGR GNSS station from the RESIF-RENAG network (http://renag.resif.fr) on the bedrock close to the glacier 3 km uphill (see yellow star in Fig. 1a). The GNSS antennas on the glacier are installed on 8-m long aluminum masts anchored 4-m deep in the ice and thus emerging about a meter above the snow surface at the beginning of the measurement period. The temporary station placed next to the glacier side provides a useful reference for validating kinematic GNSS processing approaches, evaluating station positions from every single set of GNSS signal recordings (i.e. every second, as opposed to static processing, which cumulates GNSS signals over a much longer time). We conduct such kinematic processing using the TRACK software ((Herring *et al.*, 2018), http://geoweb.mit.edu/gg/docs.php). Our processing chain includes the use of the online tool SARI (https://alvarosg.shinyapps.io/sari/) for the removal of outliers that arise from low satellite coverage in the glacier valley and to perform a de-trend and re-trend analysis to
estimate and correct for offsets due to manual antenna mast shortening as snow melt progresses. We also correct for multi-path effects induced by GNSS signal reflections from the ground, although we find that those are attenuated by the combination of GPS and GLONASS signals thanks to their different sidereal periods (~24 h for GPS and ~8 days for GLONASS). We finally calculate position time series at a 30-s time step sufficient to capture glacier dynamics and subsequently evaluate three-dimensional velocities by the linear trends of the position components. The horizontal velocity is calculated as $v_+ = -v_./ + v_1/$ where $v_.$ and $v_1$ are the North and East components, respectively.

To facilitate comparison of basal sliding and surface velocity here we smooth both timeseries at a 36-hr timescale (Fig. 3b), since daily down to sub-daily fluctuations in basal sliding velocities are largely affected by unconstrained variations in the local ice roughness in contact with the wheel, as for example when an ice-carried rock debris passes over the wheel. Although basal sliding velocity is to be lower than surface velocity, here both quantities have similar absolute values because the sliding velocity is measured at a place where the glacier is much steeper and thus driving stress is much larger than at the GNSS locations. We observe an increase in basal sliding velocity from 4.5 mm/h to more than 6 mm/h at the very beginning of the monitored period. This acceleration is not seen in the GNSS observations, which could be due to the glacier seasonal acceleration occurring earlier at this location. We also observe one major glacier acceleration event in the location of the dense seismic array occurring between May 4 and May 8 likely due to the large concomitant increase in water discharge (see Fig. 3a) causing basal water pressurization (Cuffey and Paterson, 2010).

**PRELIMINARY RESULTS**

SEISMIC NOISE CHARACTERISTICS

We investigate the spatial and temporal variability of seismic power $P$ (in dB) across a wide range of frequencies by applying Welch's method (Welch, 1967) over 4 seconds-long vertical ground motion timeseries (with 50 % overlap) prior to averaging power (in the decibel space) over 15 minutes-long time windows. This two-step strategy allows limiting the influence of impulsive events (which are studied in more details in the next sections) on the seismic power while enhancing that of the background continuous noise (Bartholomaus *et al.*, 2015; Nanni *et al.*, 2020). In Fig. 4 we present 1-100 Hz spectrograms (i.e. seismic power at any given frequency and time) over the first half of the instrumented period (April 25 to May 14) together with timeseries of 2-20 Hz frequency median seismic power at 5 different stations of the array, four of which are located on the four array sides and one of which in the array center (see node numbers in Fig. 1b and Fig. S1 for spectrograms across all stations and over the entire frequency range and experimental period). Time periods when sensors tilted as a result of snow melt causing them no longer buried are manifested by drastically reduced seismic power values across the whole frequency range (see node 6 from May 8 to May 11). Fortunately, sensor tilt only occurred at a small number of seismic stations (11 out of 98) and during a restricted time duration (less than 2 days on average, see Fig. S1). We also observe that seismic power did not change significantly from prior to after sensor reinstallation in May 11, which suggests that these are not significantly affected by potential changes in sensor coupling to snow.

All stations generally experience similar multi-day (e.g. four days' average, see black lines) variations in seismic power that are highly correlated with multi-day discharge variations (see also Fig. 3a), although seismic power precedes discharge variations by several days likely as a result of it being primarily set by the hydraulic pressure gradient that is highest during periods of rising discharge (Gimbert *et al.*, 2016; Nanni *et al.*, 2020). Although shorter term (e.g. diurnal) variations in seismic power are also similar across stations when discharge is low (from April 24 to April 28 and from May 1 to May 5) and anthropogenic noise dominates (Nanni *et al.*, 2020), the picture is different at higher discharges when seismic power is caused by subglacial water flow. In April 29 and from May 5 to May 14 seismic power exhibits pronounced (up to 10 dB) and broad frequency (1-100 Hz) short time scale (sub-diurnal to diurnal) variations at certain stations (e.g. node 6 (Fig. 4a), node 44 (Fig. 4c) and node 50 (Fig. 4d)) while not at others (e.g. node 38 (Fig. 4b) and 95 (Fig. 4e)). We also observe that at certain stations seismic power appears to be continuously or intermittently enhanced at distinct frequencies, as for instance node 38 that systematically presents much higher seismic power above 20 Hz and node 44 that presents particularly high power at frequencies around 20 Hz from April 27 to May 1. These discrepancies suggest that ground motion amplitude measurements are sensitive to a heterogenous and intermittent subglacial hydrology network and/or to subglacial water flow sources exhibiting strong directivity variations.

DETECTING AND LOCATING BASAL STICK SLIP IMPULSIVE EVENTS USING TEMPLATE MATCHING

We detect high-frequency (>50 Hz) basal stick-slip events using template matching. We follow a two-step analysis as in Helmstetter *et al.* (2015). We first build a catalog of events through applying a short-term-average over long-term-average (STA/LTA) detection method (Allen, 1978) to the continuous high-pass filtered signal (>20 Hz) and identifying an event when the STA/LTA ratio exceeds a factor of 2. We then select all events with short duration (<0.2 s) and high average frequency (>50 Hz) and define groups of events referred to as clusters when their correlation with each other exceeds 0.8. For each cluster, we compute the average waveform to define the "template" signal associated with this cluster. We visually check that events present distinct P and S wave arrivals and use a polarization analysis to ensure that they are not associated with surface waves (Fig. 5a). We then use the template matching filter method (Gibbons and Ringdal, 2006) in order to detect smaller amplitude events that are not picked but belong to the identified clusters. This analysis is conducted using the borehole station, which has a higher sensor sensitivity and sampling rate compared to the nodes.

We identify 31 active clusters during the dense array experiment period. Interestingly, these clusters constitute a large part of the 46 clusters identified on a much longer period (from December 2017 to June 2018, using the borehole sensor which ran almost continuously, see Fig. 6). Although the amplitude of these signals varies quite strongly through time (Fig. 6a), waveform characteristics strikingly remain similar (Fig. 6b). All 46 identified clusters exhibit similar characteristics to that shown in Fig. 6, and their activity does not appear to be temporally correlated with each other, nor with external drivers related to meteorology, hydrological or glacier dynamics.

We retrieve the position of the 31 identified clusters by first manually picking on each node the P and S arrival times associated with the event in each cluster that is associated with the largest correlation with the template event (see orange crosses in Fig. 5a), and then inverting for the location of each event and the associated P and S waves velocities assuming velocities are homogeneous and identical for all events. We do not consider refracted waves because the first arrival is the direct wave for most sensors and most events. Moreover, even when the refracted wave is faster, it is usually less impulsive and has a smaller amplitude than the direct wave. We look for P and S wave velocities using a grid search analysis with a step of 10 m.s$^{-1}$ and the Nonlinloc software (Lomax *et al.*, 2000) to locate clusters. We assume a standard error of arrival times of $2 \cdot 10^{-4.5}$ s for P waves, $4 \cdot 10^{-4.5}$ s for S waves and of $3.5 \cdot 10^{-4.5}$ s for calculated travel times. We can see in Fig. 5a that the picked arrival times (black circles) are in good agreement with the computed travel times (green lines). The root-mean-square error for this event is 2.4 ms, which corresponds to about one sample (2 ms).

We show the locations of basal icequakes versus depth in an average transverse section in Fig. 7a and on a two-dimensional map in Fig. 8. They are mainly located in the lower part of the array and in the central part of the glacier or in North-East side, while there is no event observed in the South-West side. Icequake depths range between 80 m and 285 m, and are in good agreement with the bedrock topography estimated from the radar profiles. Uncertainty on absolute source depth is on the order of 10 m (see errorbars in Fig. 7b), and the estimated seismic wave velocities of $V_p$=3620 m.s$^{-1}$ and $V_s$=1830 m.s$^{-1}$ (Fig. 7b) are in good agreement with velocities measured on other alpine glaciers (Podolskiy and Walter, 2016). $V_s$ is much better constrained by the data compared to $V_p$ (Fig. 7b), however the good match between icequake depth and bedrock topography suggests that our inferred seismic wave velocities correspond to reasonable estimates.

SYSTEMATIC LOCATION OF EVENTS USING MATCHED-FIELD PROCESSING

Contrary to in the previous section where a priori constraints on waveform characteristics and wave velocity are used to target basal stick-slip events, here we aim to test the capability of locating a wide range of seismic events generated by naturally occurring sources (either impulsive or continuous) with no a priori knowledge on waveform characteristics and minimal a priori knowledge on medium properties. The rationale is that the limited a-priori knowledge for source identification is balanced by the high spatial and temporal resolution provided by the array processing technique, which may enable the emergence of characteristic patterns that may be used for source identification.

We conduct Matched-Field Processing (MFP), which consists of recursively matching a synthetic field of phase delays between sensors with that obtained from observations using the Fourier transform of time-windowed data (Vandemeulebrouck *et al.*, 2013; Chmiel *et al.*, 2019). We obtain the synthetic field from a source model with a frequency-domain Green's function that depends on 4 parameters, which are the source spatial coordinates $x$, $y$ and $z$ and the medium phase speed $c$. The MFP outputs range from 0 to 1. The closer to 1, the more the modelled phase matches observations, and therefore the more likely the source model properties represent the true source properties. Here we use a spatially homogeneous velocity field within the glacier, which has the advantage of a fast-analytical computation but also results in more ambiguity between $z$ and $c$. Contrary to classical beamforming techniques in which a planar wave front is often assumed, our MFP approach considers spherical waves and allows locating sources closer to and within the array. To build a large catalog of events, we apply MFP over short time windows of 1-s with 0.5-s overlap, across 16 frequency bands of ±2 Hz width equally spaced from 5 to 20 Hz and over the entire period. Calculating source locations over such a large number of windows requires minimizing computational cost. We do so by using a minimization algorithm that relies on the downhill simplex search method (Nelder-Mead optimization) of Nelder and Mead (1965) and Lagarias *et al.* (1998) instead of using a multi-dimensional grid search approach. As the exploration of the solution space is characterized by a certain level of randomness, we maximize the likelihood that our minimization technique finds a global minima and thus the dominant source over the considered time window through (i) starting the optimized algorithm from a set of 29 points located at a depth of 250 m inside and near the array (see black crosses in Fig. 9d) with a starting velocity $c$=1800 m.s$^{-1}$ and (ii) taking the highest MFP output out of the 29 inversions found after convergence.

In Fig. 9b,c we present two examples of events located inside and outside the array and associated with a high MFP output of 0.92. The half-size of the focal spot in the MFP output field gives a measure of the location uncertainty (Rost and Thomas, 2002), which is about 10 m for events located inside the array and can increase up to 40 m when for events up to 100 m away from the array edges. Gathering all sources over one continuous day of record, we find that the associated MFP outputs distribution exhibits a heavy tail towards high values (see red area in Fig. 9a for an example at 13 Hz). Such a heavy tail is not obtained for a random field, in which case MFP outputs exhibit a distribution shifted towards almost one order of magnitude lower values. This suggests that most identified sources correspond to real and detectable seismic events. Well resolved seismic events with MFP outputs higher than 0.8 are located near the surface and nicely delineate crevasse geometries, such that they likely correspond to englacial fracturing (see red dots in Fig. 8). A restricted number of these events are however located outside of the glacier and likely correspond to rock falls. Typical waveforms associated with englacial fracturing events exhibit clear P and surface waves arrivals (Fig. 5b), as well as hyperbola arrivals that likely correspond to reflected waves at the glacier/bedrock interface (see black arrows in Fig. 5b).

USING CATALOGS OF EVENTS FOR STRUCTURE INVERSION

Dense-array techniques for seismic imaging often involve interferometry analysis on continuous seismic noise. Such techniques however require an equipartitioned wavefield inherited directly from homogenously distributed noise sources and/or indirectly from sufficiently strong scattering (Lobkis and Weaver, 2001; Fichtner *et al.*, 2019). These conditions strongly limit the applicability of such techniques on glaciers where sources are often localized and waves in ice being weakly scattered (Sergeant *et al.*, 2020). An alternative way is to use localized and short-lived sources with known positions (Walter *et al.*, 2015) as those previously identified using our systematic MFP technique, which are numerous and quite evenly distributed in space (Fig. 8).

We consider the catalog of sources associated with MFP outputs larger than 0.6, located near the surface (z<10m) and close to the array (within a radius of 400 m from the array center). With these criteria our catalog includes about $10^6$ sources gathered over the 35 days of continuous recordings. In order to further demonstrate that our MFP output inversions yield reliable velocities (i.e. the ambiguity between $z$ and $c$ is limited for these sources), we use the velocities as outputted from our MFP algorithm to establish the observed dispersion curve, as opposed to conducting a classical f-k analysis (Capon, 1969). We infer surface wave phase velocity at each frequency between 3.5 Hz and 25 Hz by fitting a Gaussian function to the probability density distributions of velocities in each frequency bin, and taking the center of the Gaussian function as the most representative velocity in that frequency bin (see Fig. 10a (inset) for an example at 13 Hz). We note that the presently constructed dispersion curve is similar to the one that would be obtained using a classical f-k analysis (not shown). We find that surface wave velocity increases gently from 1560 m.s$^{-1}$ to 1630 m.s$^{-1}$ as frequency decreases from 25 Hz down to 7 Hz, and then increases sharply up to 2300 m.s$^{-1}$ as frequency decreases down to 3.5 Hz. These observations can be reproduced using a three-layer onedimensional elastic model (using the Geopsy package, see Wathelet *et al.* (2020)) that incorporates a gentle velocity increase (from 1670 to 1720 m.s$^{-1}$ for $V_?$) at 40 m depth and a drastic velocity increase (from 1720 to 2800 m.s$^{-1}$ for $V_?$) located between 200 and 220 m depth (Fig. 10b). These values were obtained by trial and error tests. The slightly slower velocities and density within the first 40-m deep layer may be due to surface crevasses, and are consistent with surface events being associated with smaller P wave velocities than those associated with stick-slip events at the ice/bedrock interface (Fig. 5). The 200- to 220-m deep drastic discontinuity reflects the ice/bedrock interface, consistent with the radar-derived average glacier thickness beneath the seismic network (Fig. 2a).

We go one step further and perform two-dimensional surface wave inversions from eikonal wave tomography (Roux *et al.*, 2011; Lin *et al.*, 2013; Mordret *et al.*, 2013). We first extract ~200,000 Rayleigh wave travel times using the best (associated with MFP outputs larger than 0.9) seismic events and then perform a simple linear inversion for the slowness (starting from a homogeneous initial model with a phase velocity of 1580 m.s$^{-1}$, see Fig. 10a) assuming straight rays as propagation paths and an *a-priori* error covariance matrix that decreases exponentially with distance over 10 m. The weight of the spatial smoothing is chosen at the maximum curvature of the standard trade-off analysis (L-curve) based on the misfit value (Hansen and O'Leary, 1993), and the inversion produces a residual variance reduction of ~98% relative to the arrival times for the homogeneous model. In Fig. 8 we show the Rayleigh wave phase velocity maps obtained as a result of the travel-time inversion on a regular horizontal grid with steps of 5 m and using 13-Hz Rayleigh waves, which have largest sensitivity between 20 m and 60 m depth (Fig. 10c) according to kernel sensitivity computations performed on the three layer elastic model (Fig. 10b) using the code of Herrmann (2013). We observe that locations with higher crevasses density are generally associated with lower phase velocities, as observed in the left and bottom sides of the array. This observation is however not systematic, since high velocities are also observed in the top right and top side of the array where crevasses are also present. This could be explained by crevasses being shallower or by crevasses having a different azimuthal orientation at these locations. This latter potential source of bias could be investigated by explicitly accounting for anisotropy in the tomography inversion scheme (Mordret *et al.*, 2013).

**DISCUSSION**

INTERPRETING SPATIAL AND TEMPORAL VARIATIONS IN GROUND MOTION AMPLITUDES

Although our seismic array observations generally exhibit spatially homogenous multi-day changes in seismic power, there exists specific times when changes in seismic power are spatially heterogeneous. A surprising observation is that these heterogeneous changes are observed down to the lowest frequencies (3 to 10 Hz) associated with wavelengths larger than the inter-station spacing, such that the observed spatial heterogeneity is unlikely solely caused by wave attenuation. It remains to be investigated as to which processes mainly cause the observed spatial variability in signal amplitude. Punctual sources identified from the MFP analysis could be used to investigate the respective control of wave attenuation, wave scattering and site effects on amplitude field heterogeneity and its potential dependency on site attributes like crevasse density, glacier thickness or snow layer thickness. Full waveform modelling combined with wave polarity analysis could also be conducted in order to further understand how wave focusing in the near field domain as well as source heterogeneity and directivity may cause heterogenous amplitude wavefields. Incorporating these constraints into an improved model describing the control of both source and wave propagation physics on the seismic wave amplitude field (Gimbert *et al.*, 2016) could allow using our dense array observations to infer the spatial variability in subglacial flow parameters such as subglacial channel size and pressure.

BETTER UNDERSTANDING THE PHYSICS OF STICK SLIP EVENTS

The application of template matching to our dense seismic array observations confirms that stick-slip events operate at the ice-bed interface, as previously suggested by Helmstetter *et al.* (2015) based on single station observations. The additional observation that events are all located in the lower part of the array and in the central part of the glacier or in North-East side, while there is no event observed in the South-West side, provides further observational support that specific bed conditions (e.g. water pressure, bed shear stress, bed roughness, bed topography, carried sediments) are necessary for these events to occur (Zoet *et al.*, 2013; Lipovsky *et al.*, 2019). Further insights into the physics controlling the spatio-temporal dynamics of these events could be gained by improving the detection scheme using all sensors instead of the borehole sensor only as presently done and performing relative event location within each cluster using double-differences methods (Waldhauser and Ellsworth, 2000) instead of simply inferring single cluster locations as presently done. These improvements could allow detecting more clusters and identifying whether or not stick-slip asperities migrate.

USING SYSTEMATIC SOURCE LOCATION TO RETRIEVE SOURCES AND STRUCTURAL PROPERTIES

Systematic MFP analysis with adequate parametrization opens a route to continuous, automatic, and statistics-based monitoring of glaciers. A wide diversity of seismic sources may be identified and studied separately with this technique by scanning through the different ranges of MFP outputs. High MFP output observations may be used to study the dynamics of crevasse propagation with particularly high spatio-temporal resolution. Such observations may allow better understanding the underlying mechanisms associated with crack propagation and in particular its modulation by water. Lower MFP outputs may be used to locate spatially distributed sources generating coherent signals only over a restricted number of array stations. These distributed sources may include tremor sources (e.g. water flow) or various glacier features (e.g. crevasses, englacial conduits) acting as scatterers. One could also combine MFP with eigenspectral decomposition to reveal weaker noise sources that would otherwise be hidden behind dominant noise sources (Seydoux *et al.*, 2016). Additional constraints for seismic imaging may also be provided through identifying specific events generating waves of particular interest for structural analysis such as in particular bedrefracted waves as shown in Fig. 5 (see black arrows).

## SUMMARY

We present a dense seismic array experiment made of 98 3-component seismic stations continuously recording during 35 days in early spring on the Argentière Glacier, French Alps. The seismic dataset is supplemented by a wide range of complementary observations obtained from ground penetrating radar, drone imagery, GNSS positioning and in-situ instrumentation of basal glacier sliding velocities and subglacial water flow discharge. We show that a wide range of glacier sources and structure characteristics can be extracted with high definition through conducting multiple seismic processing techniques including event detection from template matching and systematic matched-field processing. Future studies focusing more specifically on each aspect of the herein presented observations may enable yielding novel quantitative insights on spatio-temporal changes in glacier dynamics and structure.

## DATA AND RESOURCES

Raw seismic data can be found at:

Roux, P., Gimbert, F., & RESIF. (2021). *Dense nodal seismic array temporary experiment on Alpine Glacier of Argentière (RESIF-SISMOB)* [Data set]. RESIF - Réseau Sismologique et géodésique Français. https://doi.org/10.15778/RESIF.ZO2018 (see also link http://seismology.resif.fr/#NetworkConsultPlace:ZO%5B2018-01-01T00:00:00_2018-1231T23:59:59%5D).

Processed data used in this paper can be found at:
- https://doi.org/10.5281/zenodo.3701519 for meteorological, subglacial water flow discharge and glacier sliding speed data
- https://doi.org/10.5281/zenodo.3971815 for bed thickness, surface elevation, nodes positions, crevasses positions, surface velocity, noise PSDs, event occurrences and locations derived from template matching for stick-slip events and MFP for englacial fracturing events
- https://doi.org/10.5281/zenodo.3556552 for drone orthophotos

**ACKNOWLEDGEMENTS**

This work has been supported by a grant from Labex OSUG (Investissements d'avenir – ANR10 LABX56). IGE and IsTerre laboratories are part of Labex OSUG (ANR10 LABX56). Complementary funding sources have also been provided for instrumentation by the French "GLACIOCLIM (Les GLACIers comme Observatoire du CLIMat)" organization and by l'Agence Nationale de la recherche through the SAUSSURE (https://saussure.osug.fr, ANR-18-CE010015) and SEISMORIV (ANR-17-CE01-0008) projects. We thank C. Aubert, A. Colombi, L. Moreau, L. Ott, I. Pondaven, B. Vial, L. Mercier, O. Coutant, L. Baillet, M. Lott, E. LeMeur, L. Piard, S. Escalle, V. Rameseyer, A. Palanstjin, A. Werhlé and B. Urruty for their help in the field, as well as Martin, Fabien and Christophe for mountain guiding the group.

Figure

**FIGURES**

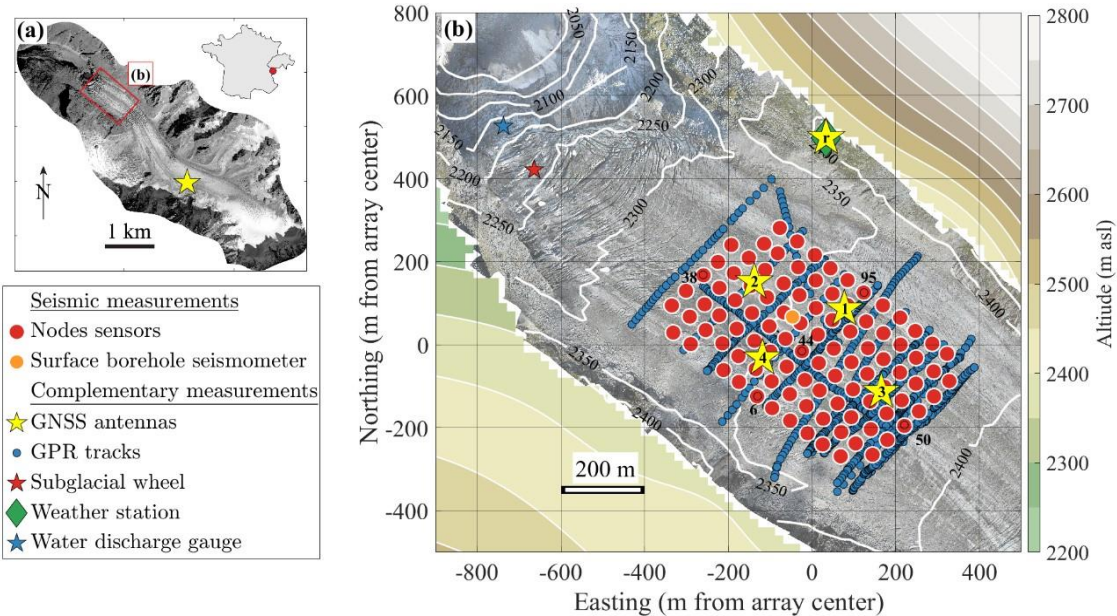

Fig. 1: Maps of the Argentière Glacier and of the instruments deployed during the dense array 5 experiment. (a) Aerial picture of the Argentière Glacier taken in 2003. The red rectangle indicates the area shown in Fig. 1(b), which we focus on in this study. The yellow star refers to a permanent GNSS station and the red dot in the inset shows the location of the glacier with 8 respect to French borders. (b) Map showing the lower part of the Argentière Glacier along with the instrument's positions. White contours indicate glacier surface topography as retrieved from structure from motion, and color contours indicate topography outside of the glacier. The 11 various symbols refer to instruments as specified in the legend. Numbers associated with red 12 circles indicate nodes that are used for illustrative examples in Fig. 4.

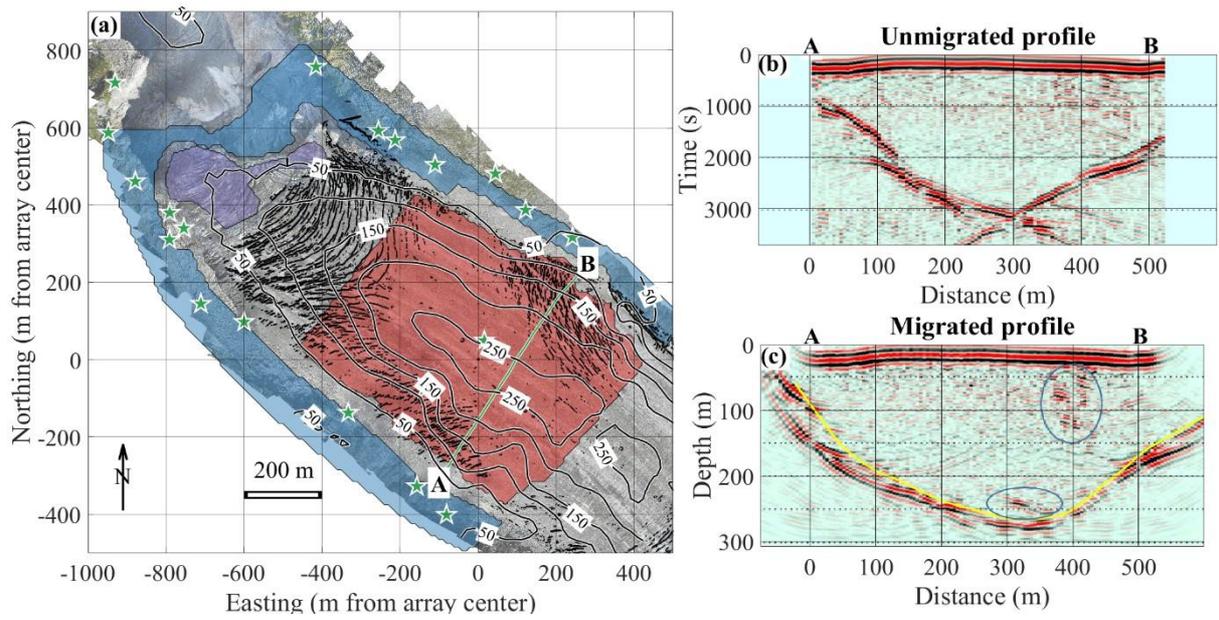

Fig. 2: (a) Reconstructed ice thickness (black contours) and surface crevasse maps (black dots), 18 along with the locations of various data constraints used to establish the surface and bed DEMs.

Green stars correspond to the GNSS measured ground control points, while colored areas differentiate between observations used to constrain the bed DEM: the blue area is from a 2018 surface DEM, the purple area corresponds to where ice-bed coordinates are known from in-situ borehole measurements and from excavated tunnels, and the red area corresponds to where glacier depth is inferred from the GPR measurements. The green line shows the track associated with the selected GPR profile shown in (b) and (c). (b) and (c) Examples of processed (b) unmigrated and (c) migrated GPR data acquired along the AB profile shown in (a). The yellow curve corresponds to the picked interface and the blue ellipses highlight local reflectivity anomalies.

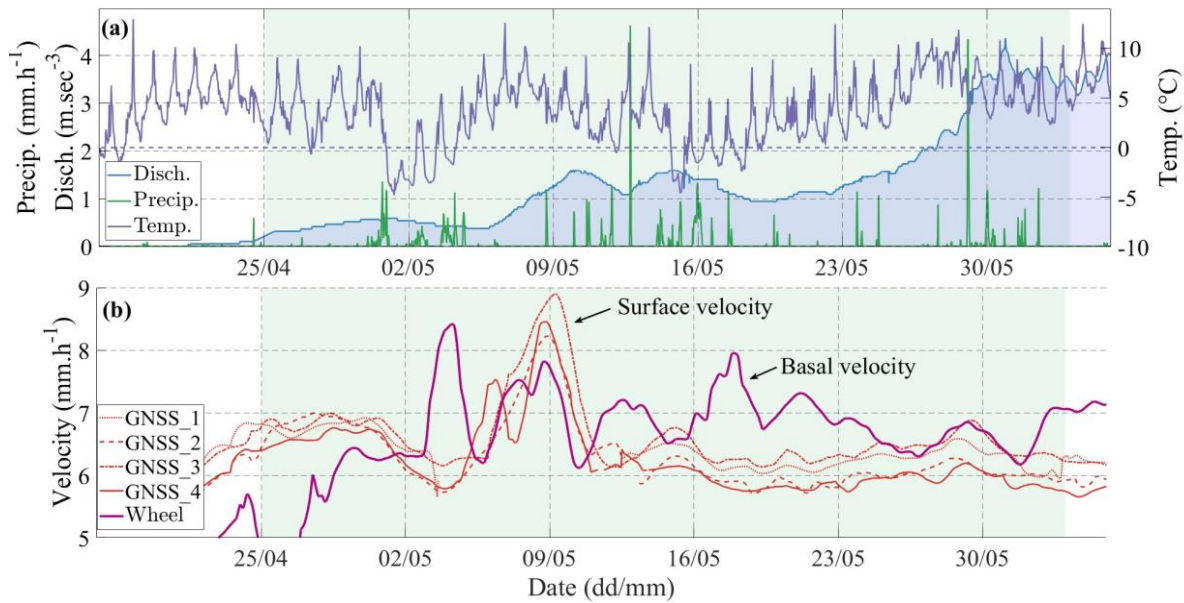

Fig. 3: Time series of physical parameters associated with meteorology, hydrology and glacier dynamics during the dense-array experiment (from April 25 to June 6, green area). (a) Glacier outlet water discharge (blue), surface temperature (purple) and precipitation (green). (b) Horizontal glacier flow velocities as measured at the glacier surface through GNSS monitoring (orange lines) and at the glacier base through direct wheel monitoring (thick purple line). See (Nanni *et al.*, 2020) for longer time series (over the 2016-2018 period).

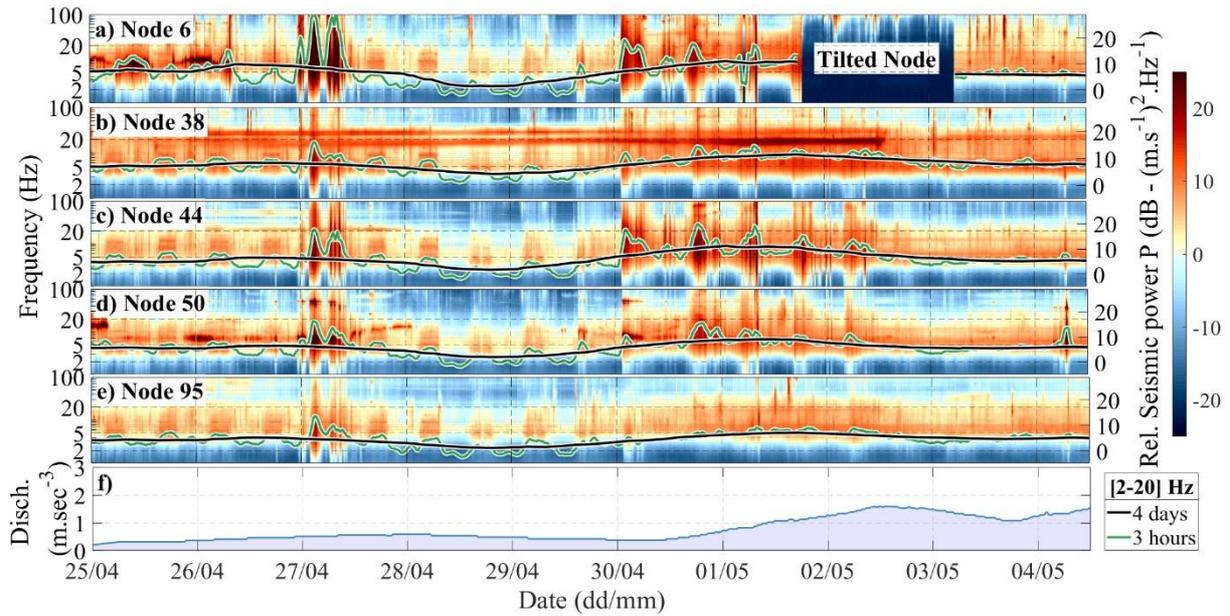

Fig. 4: (a-e) Spectrograms calculated at five selected stations (see corresponding numbers in Fig. 1) across the array from April 25 to May 13. Curves indicate 2-20 Hz frequency mean seismic power as smoothed over short (3 hours, green lines) and long (4 days, black lines) periods. See Fig. S1 for spectrograms over the whole period and across all stations. (f) Glacier outlet water discharge as shown in Fig. 3.

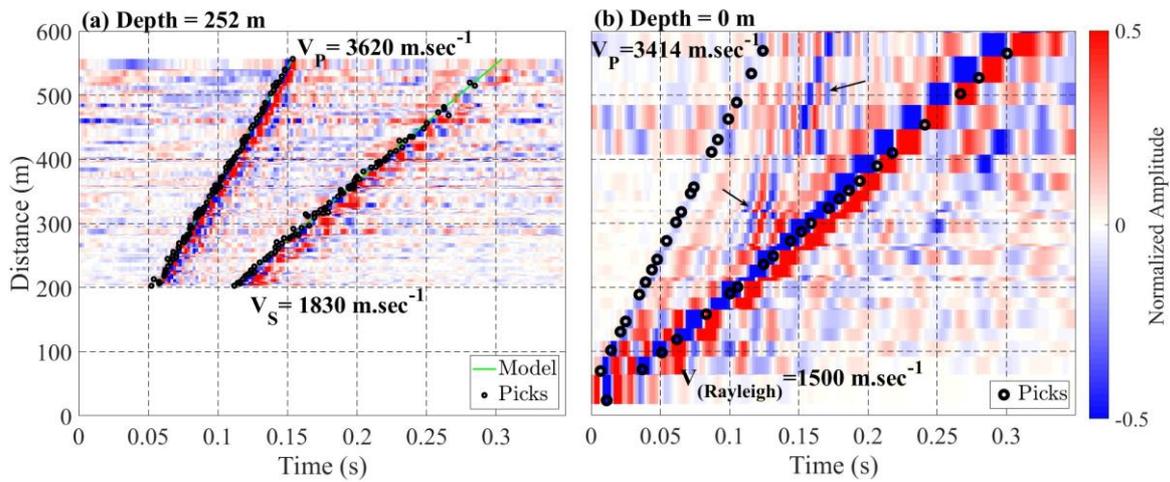

Fig. 5: Broad band seismograms of (a) a basal event as identified from template matching and (b) a surface event as identified from match-field-processing. Corresponding event locations are shown in Fig. 8. Black circles correspond to picked P, S and Rayleigh arrival times and green lines on (a) correspond to predicted arrival times using a P-wave velocity of 3620 m.sec$^{-1}$ and a S-wave velocity of 1830 m.sec$^{-1}$. A hyperbola event is also visible at large offsets on panel b (see black arrows). The zero time corresponds to the event time.

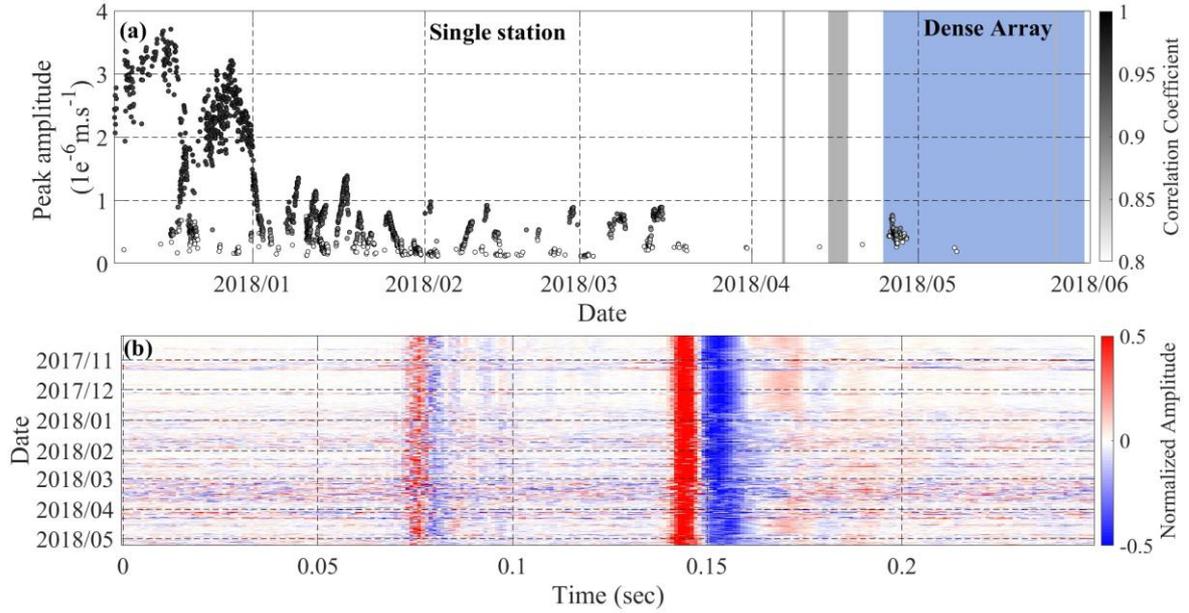

Fig. 6: a) Time series of peak amplitude for one cluster of repeating basal events. The grey scale indicates correlation with the template signal. Grey areas indicate gaps in the data and the blue area highlights the time period spanned by our dense-array experiment. b) Waveforms of all events of a cluster normalized by peak amplitude (using the North component of the borehole station). The colorbar indicates normalized waveform amplitude. Each horizontal line represents one event. The zero time corresponds to the event time.

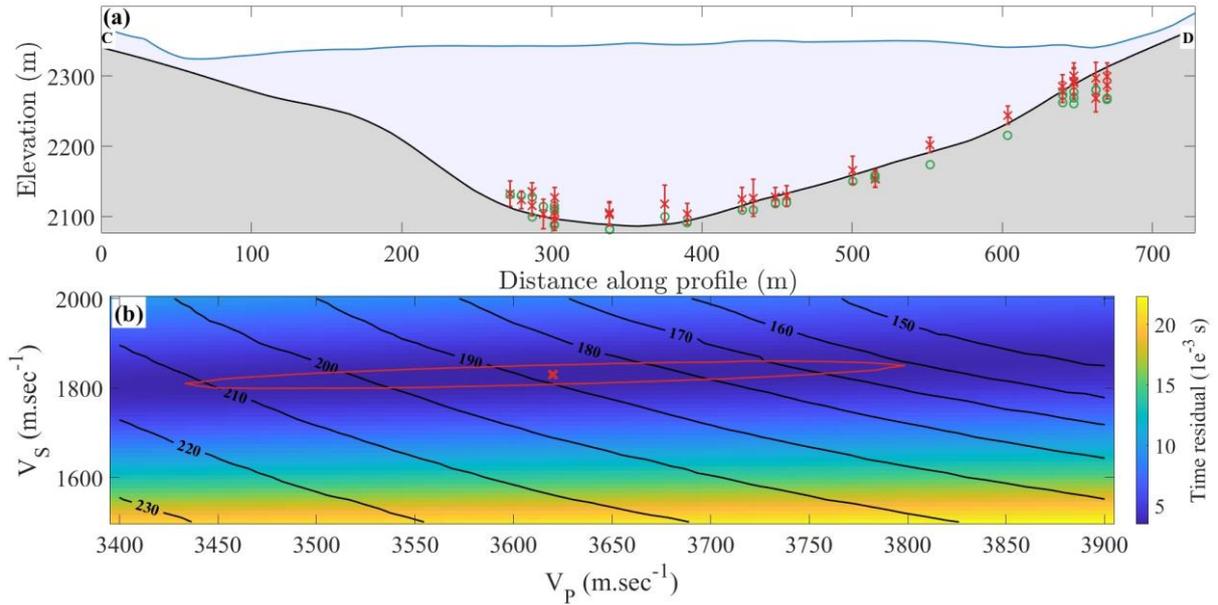

Fig. 7: (a) Two-dimensional representation of stick-slip event locations (red crosses). Red error bars show the 95% confidence interval. Green circles indicate the projected depth at the exact location of each event, while the glacier cross-section corresponds to that along the CD profile shown in Fig. 8. (b) Average time residuals (background image) and average icequake depth (black contours) as a function of the seismic wave velocities $V_P$ and $V_S$ used to locate basal icequakes. The red cross indicates the velocities $V_P$=3620 m.sec$^{-1}$ and $V_S$=1830 m.sec$^{-1}$ that minimize the average time residuals. The red line delineates the range of $V_P$ and $V_S$ associated with an average residual that is smaller than 105% of the minimum value.

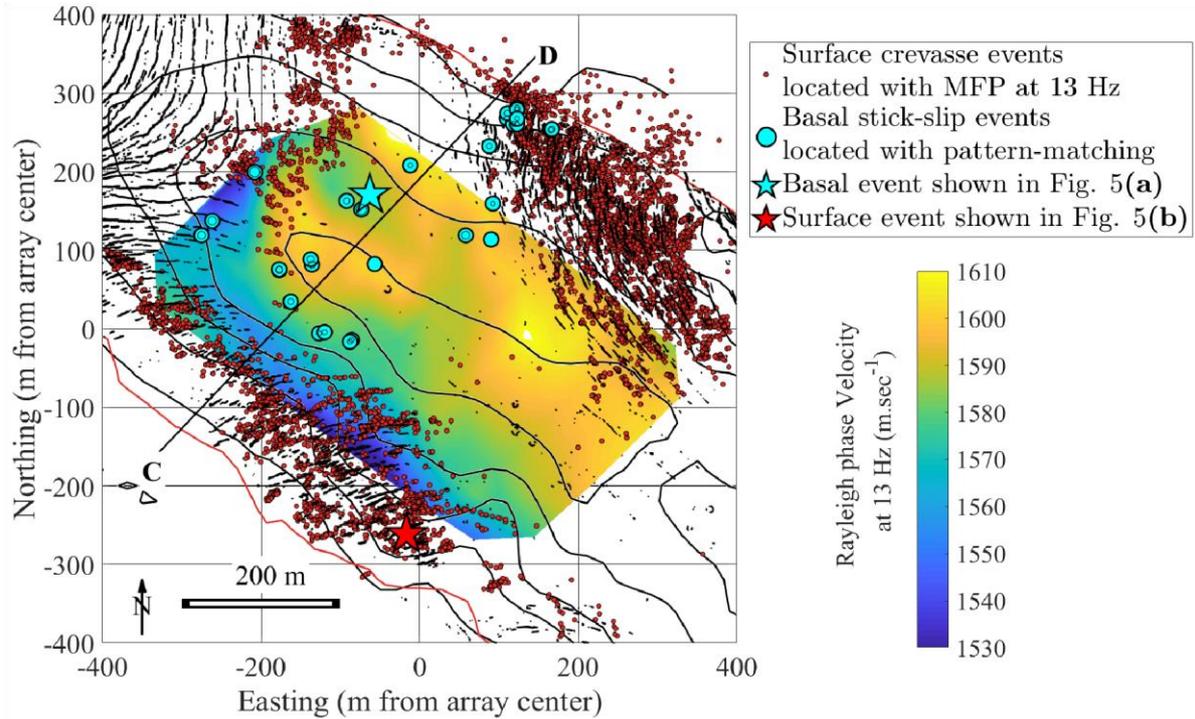

Fig. 8: Map showing the positions of basal stick-slip clusters (filled blue circles) as inferred from template matching and of icequake events located with match-field-processing at 13 Hz during the whole period with an MFP output higher than 0.8 (filled red circles). The colored area shows phase velocities from Rayleigh-wave travel-time tomography at 13 Hz. The C-D profile refers to the profile used in Fig. 7 and the blue and red stars respectively refer to the events shown in Fig. 5(a) and 5(b). Black dots show crevasses, contour lines show ice thickness (m) and the red lines delineate the glacier extent.

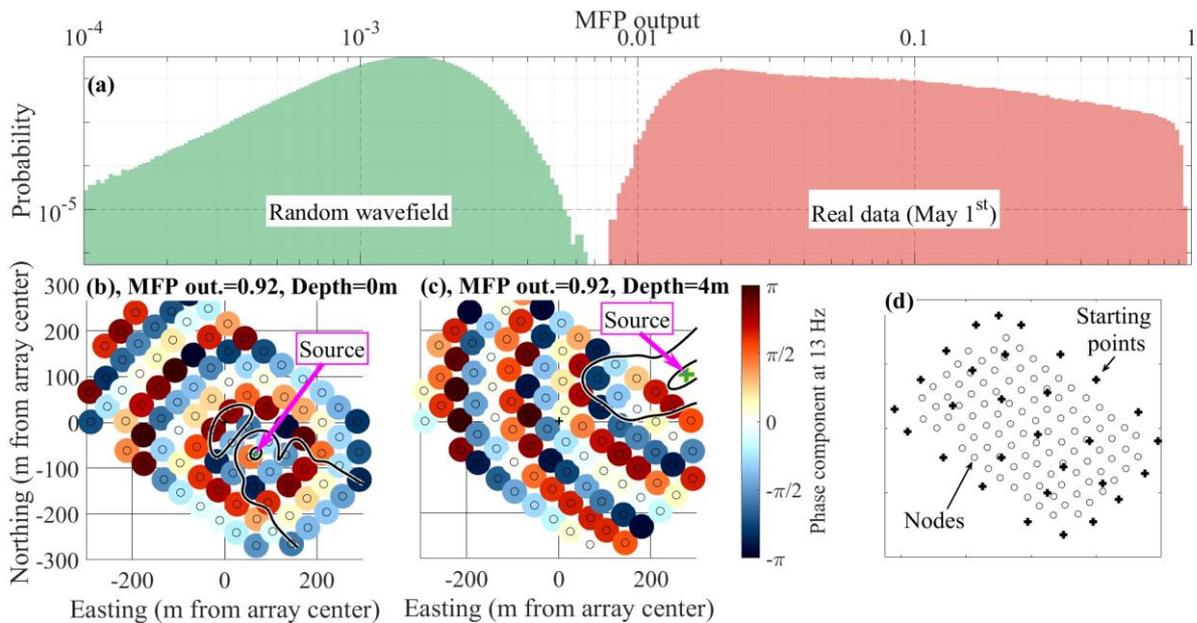

Fig. 9: (a) Distributions of MFP output values obtained at 13 Hz when applying MFP on one day (May 1) of real data (red) and on a numerically-generated random wavefield (green). The bottom panels (b) and (c) show the phase fields observed over a 1-s time window at 13 Hz for two selected events. Locations obtained from MFP using our minimization process are shown by the pink arrow/green crosses, while the contour lines show 0.1 and 0.8 MFP outputs isocontours calculated by applying grid search over the glacier surface. Panel (d) shows the locations of the 29 starting points (black crosses) used for the MFP along with nodes position (black circles).

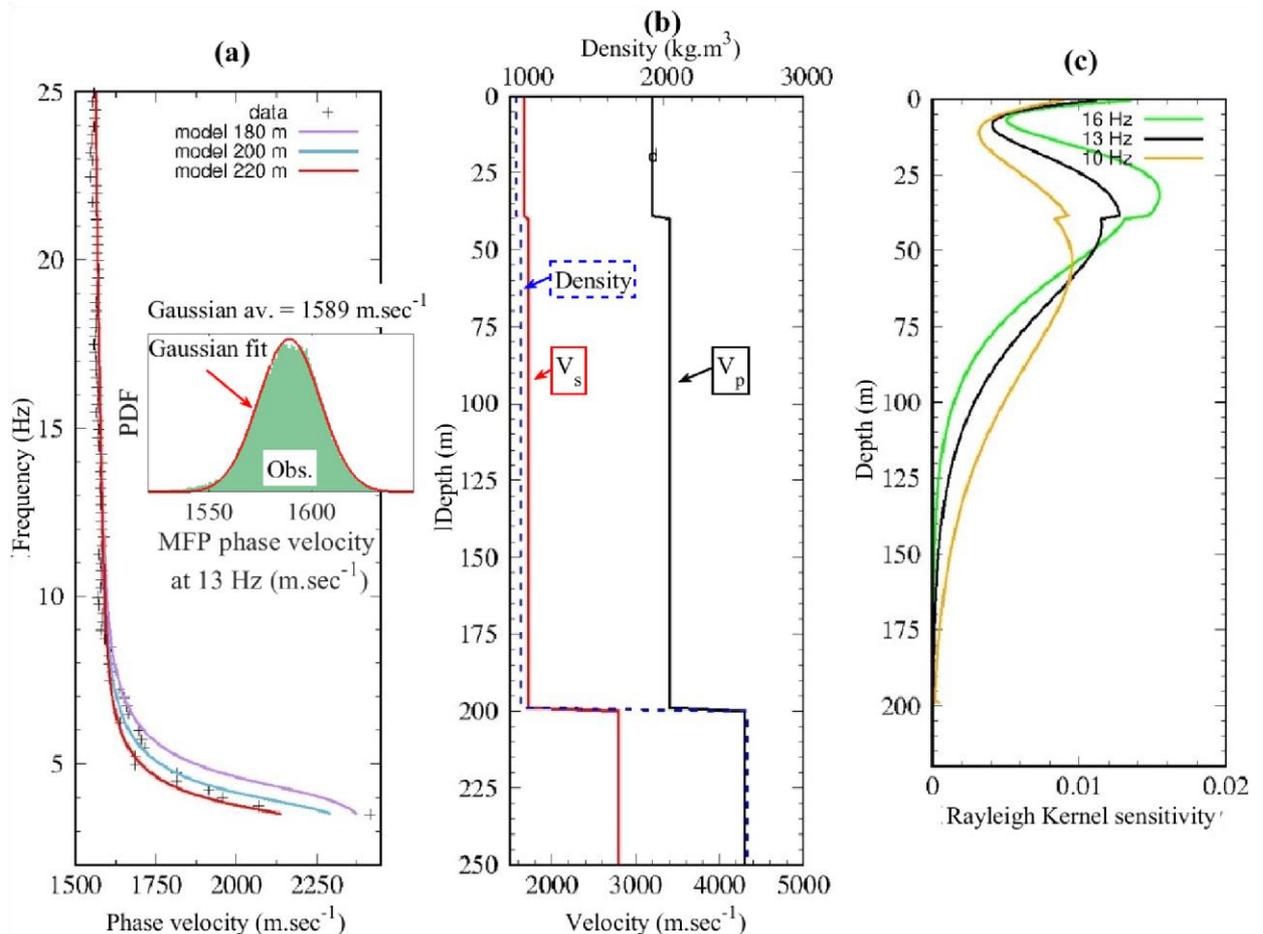

Fig. 10. Inversion of a one-dimensional structure using an average surface wave dispersion curve. (a) Comparison between the observationally-derived dispersion curve (black crosses) and synthetic Rayleigh wave dispersion curves computed using the elastic model displayed in (b) using glacier thicknesses of 180 m (purple), 200 m (blue) or 220 m (red). The inset shows the distribution of phase velocity obtained from match-field-processing at 13Hz (green) along with a Gaussian fit (red). The central value of the gaussian fit is used to establish the dispersion curve. (b) Synthetic model used to predict the observed dispersion curve. (c) Sensitivity kernels of Rayleigh waves as a function of depth for three frequencies: 16 Hz (green), 13 Hz (black) and 10 Hz (orange) associated with the glacier model shown in (b).

Supplemental Figure S1

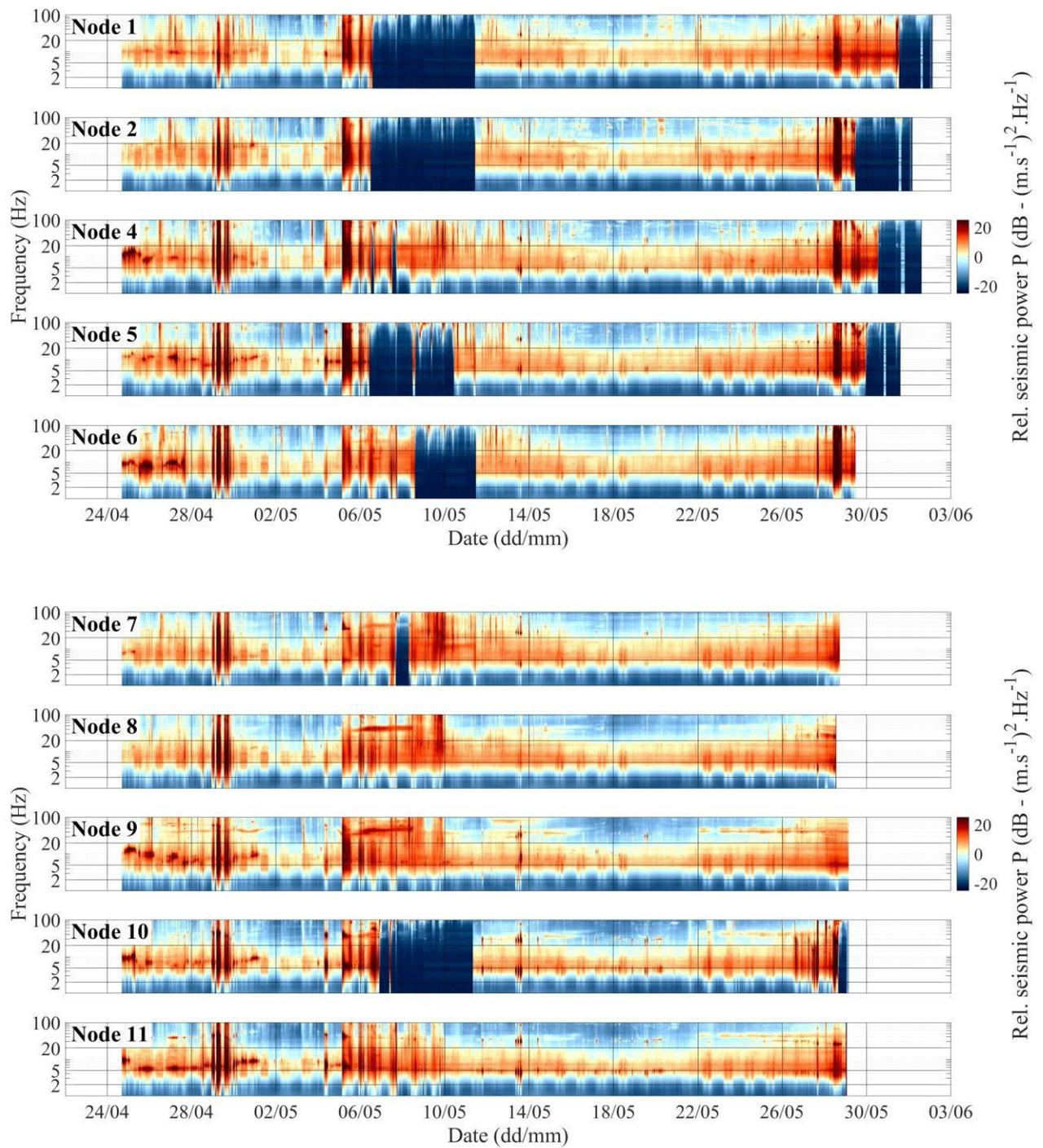

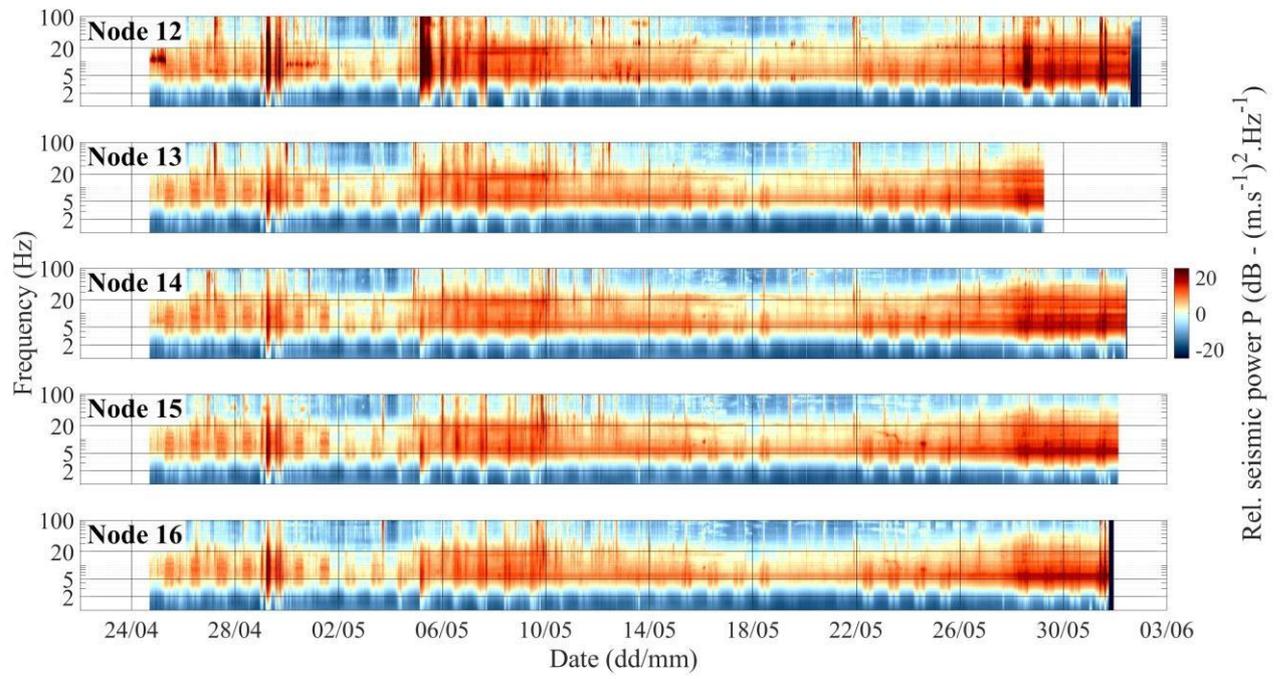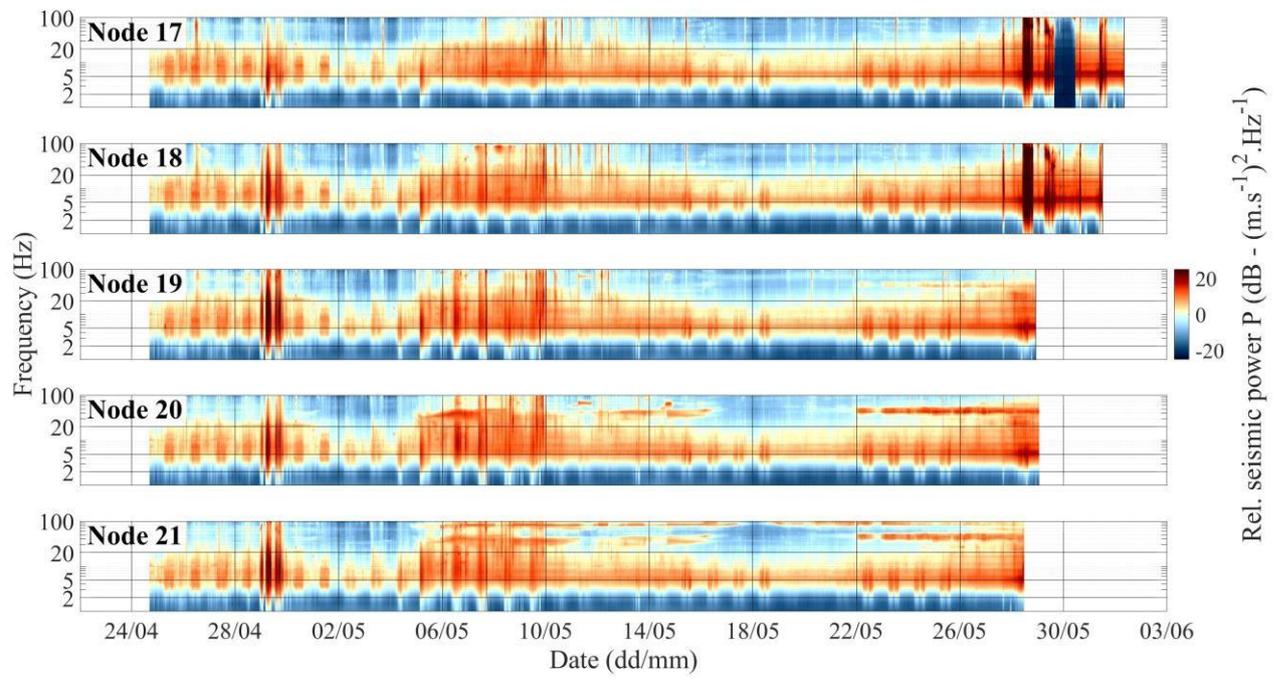

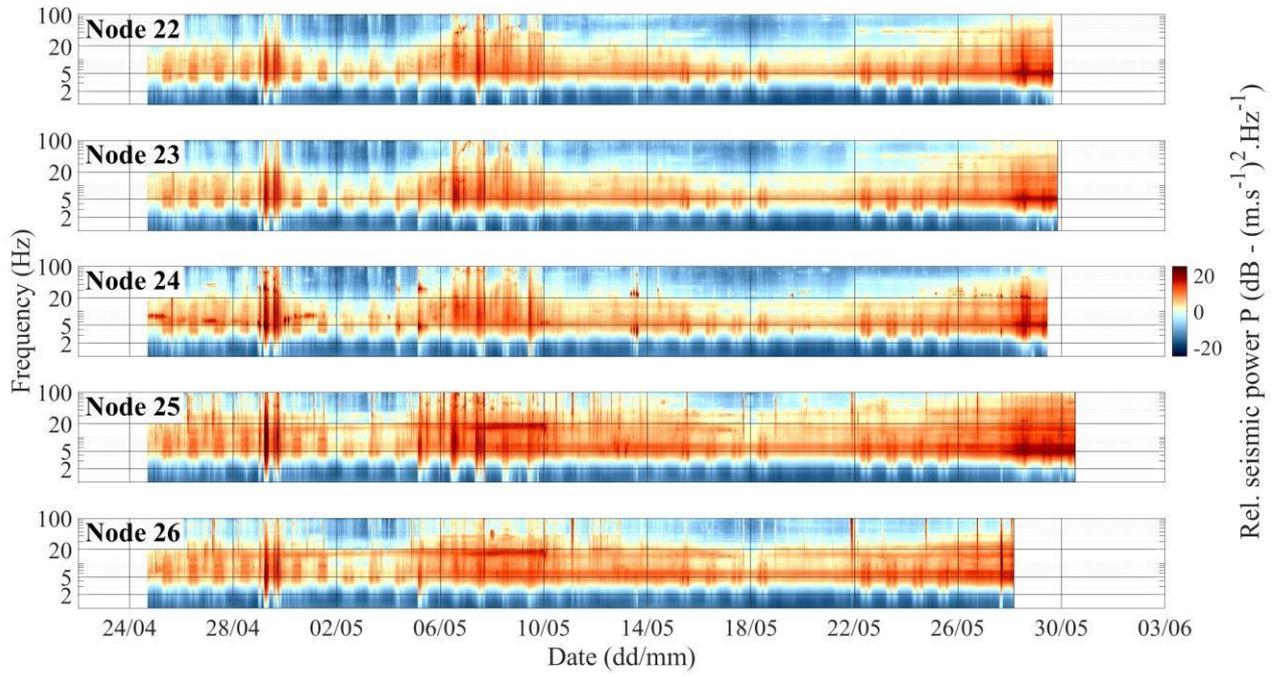
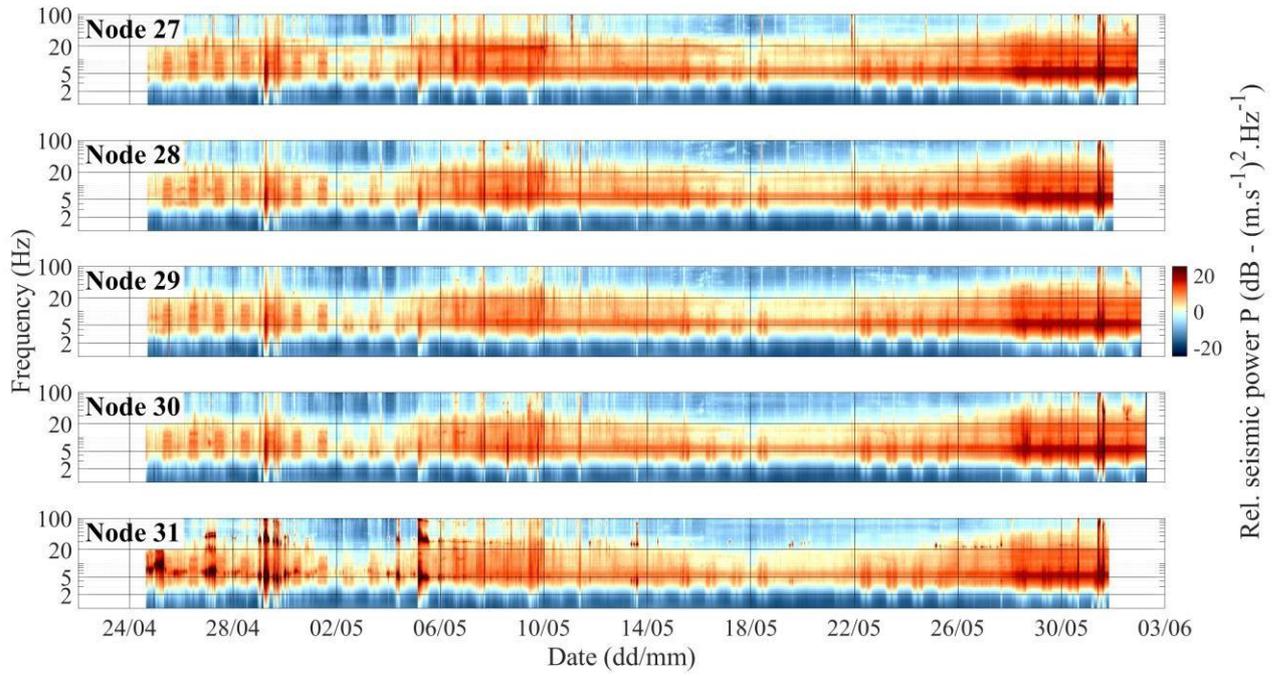

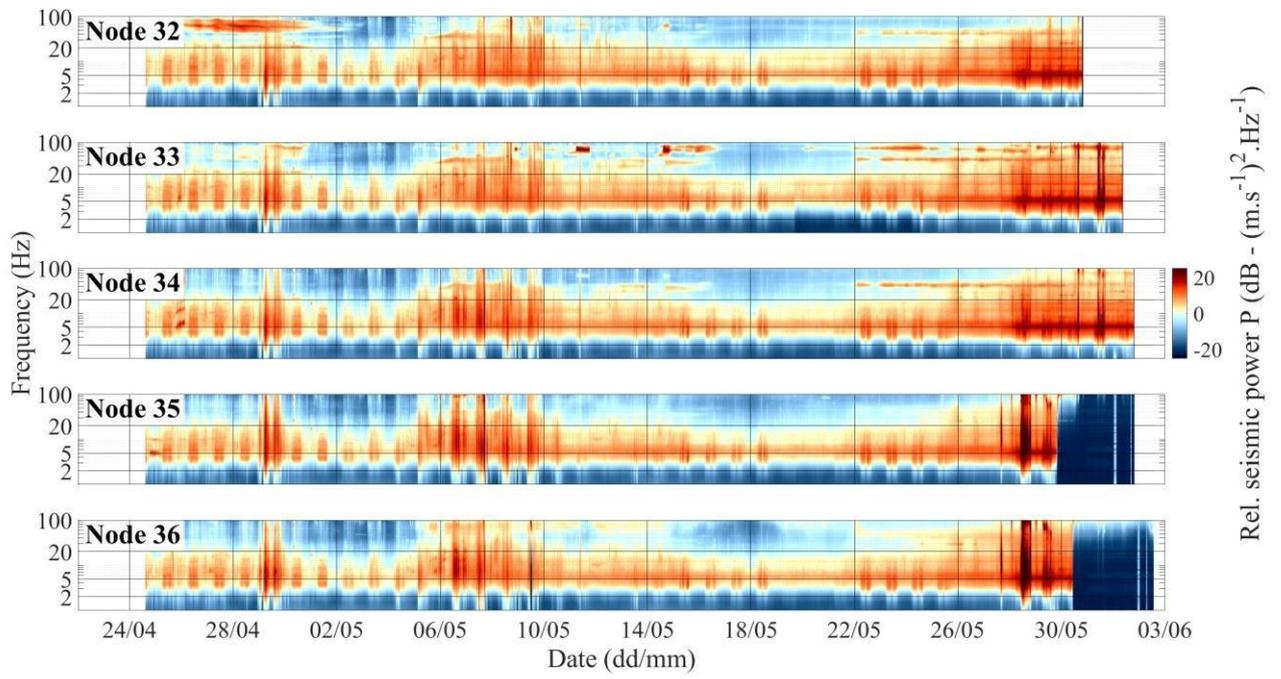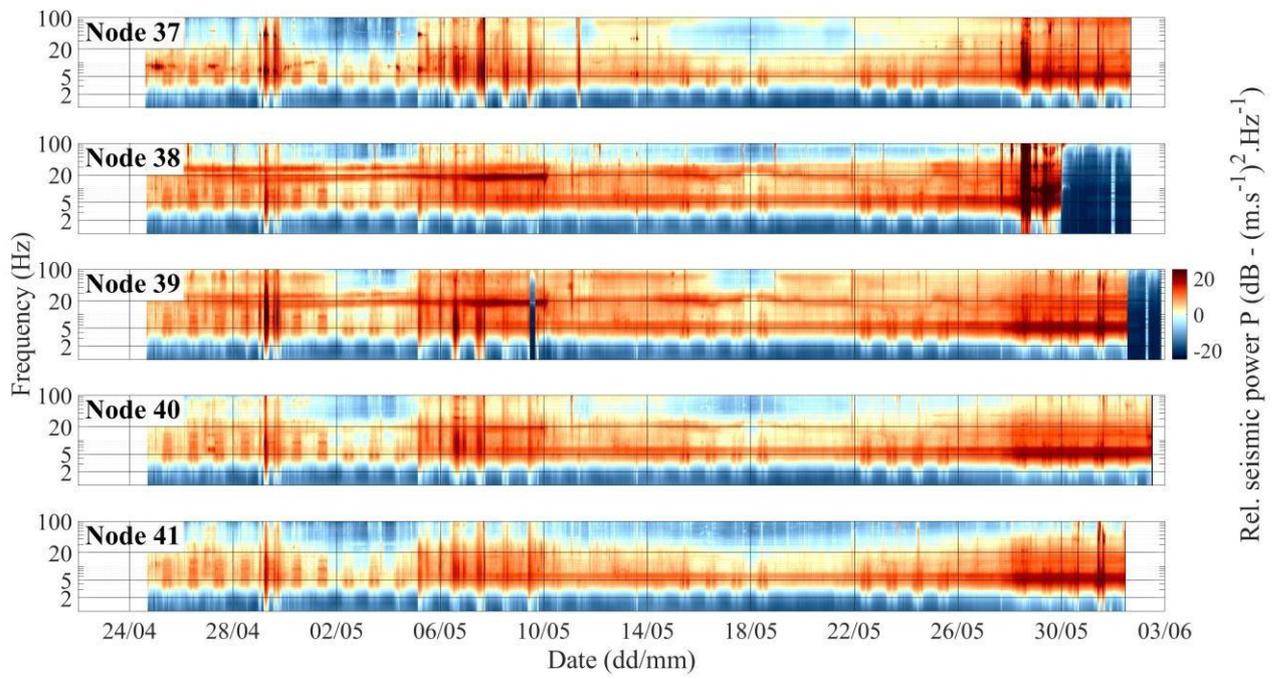

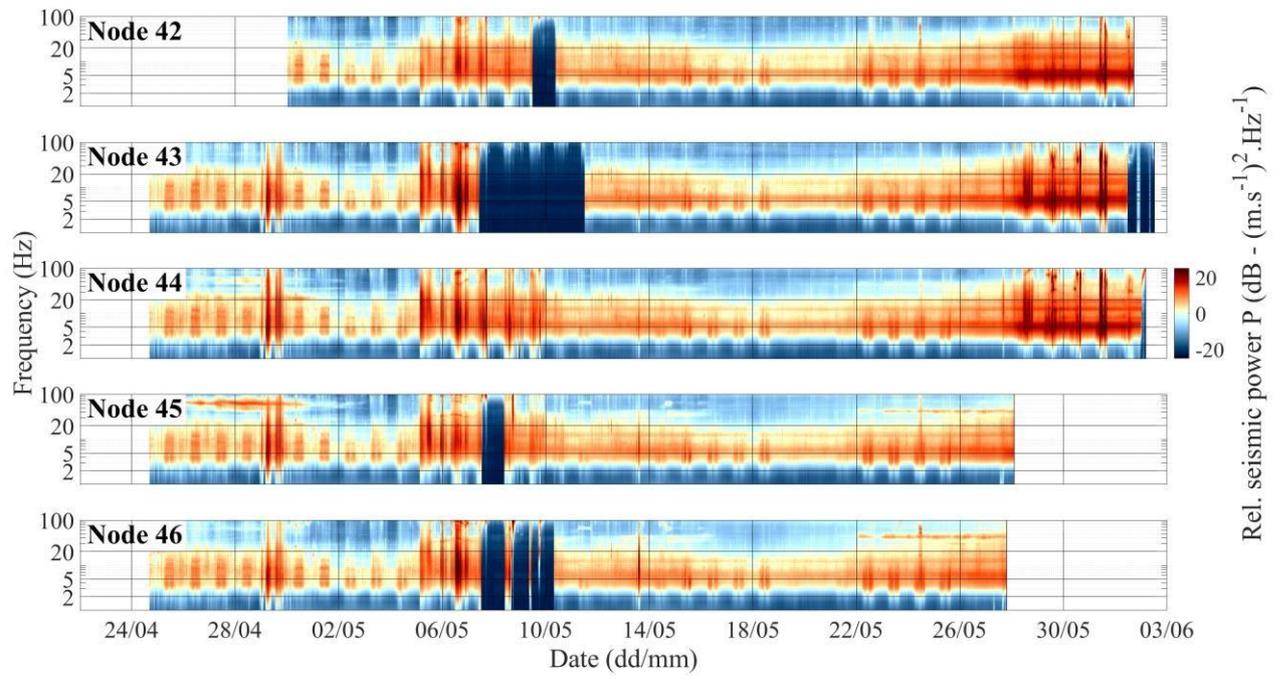
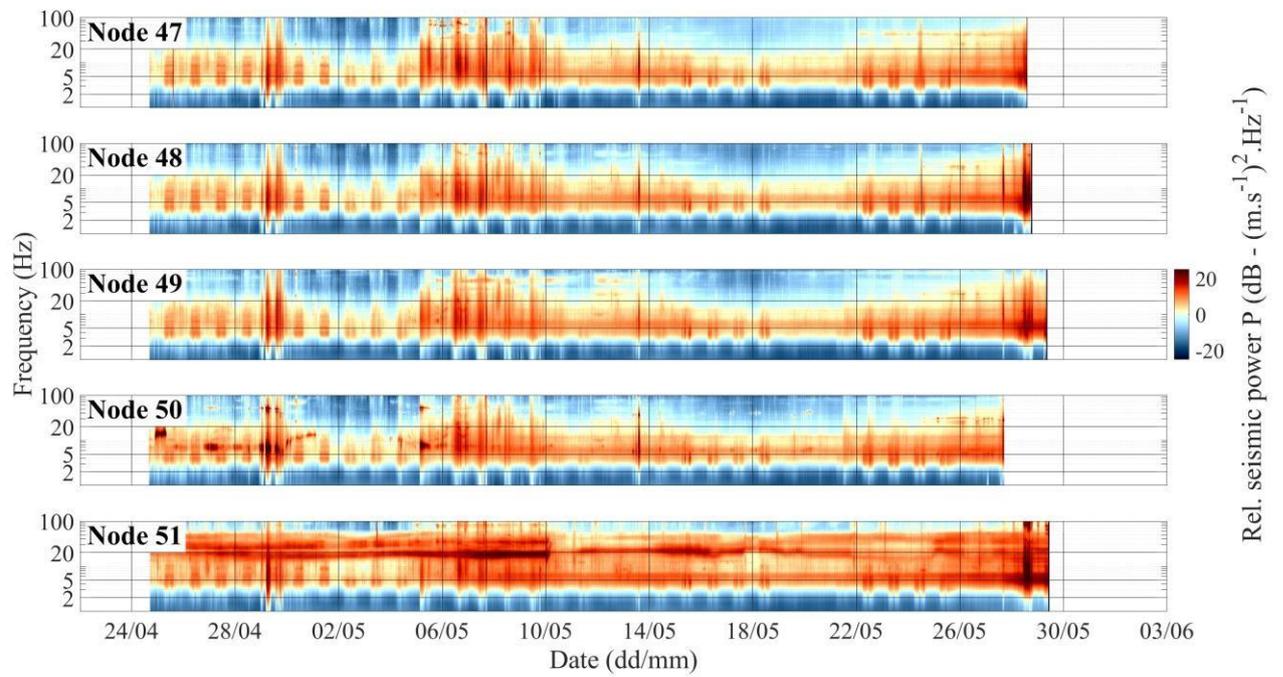

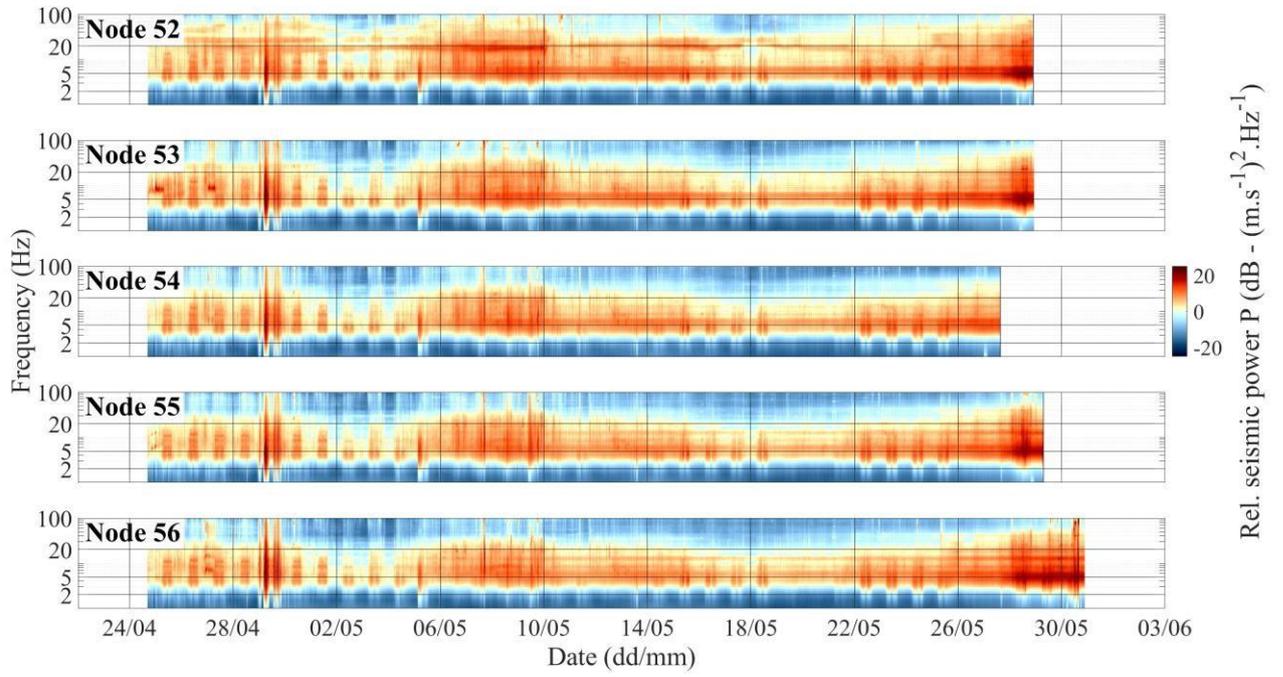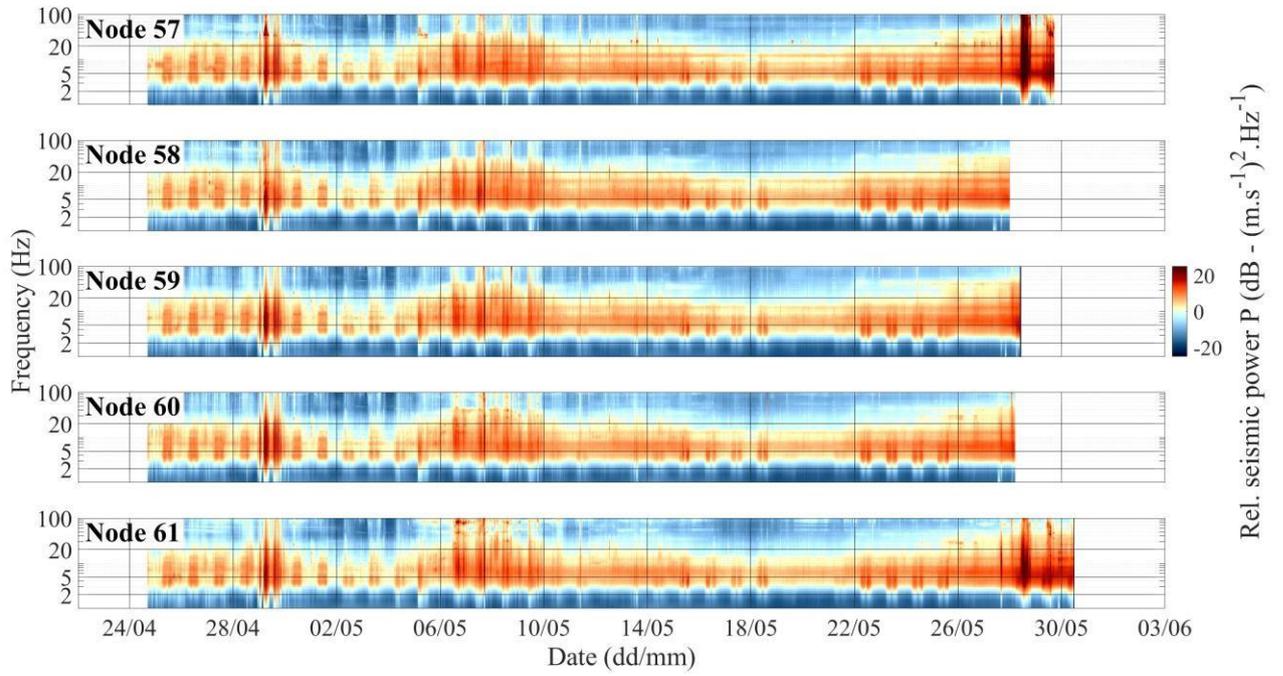

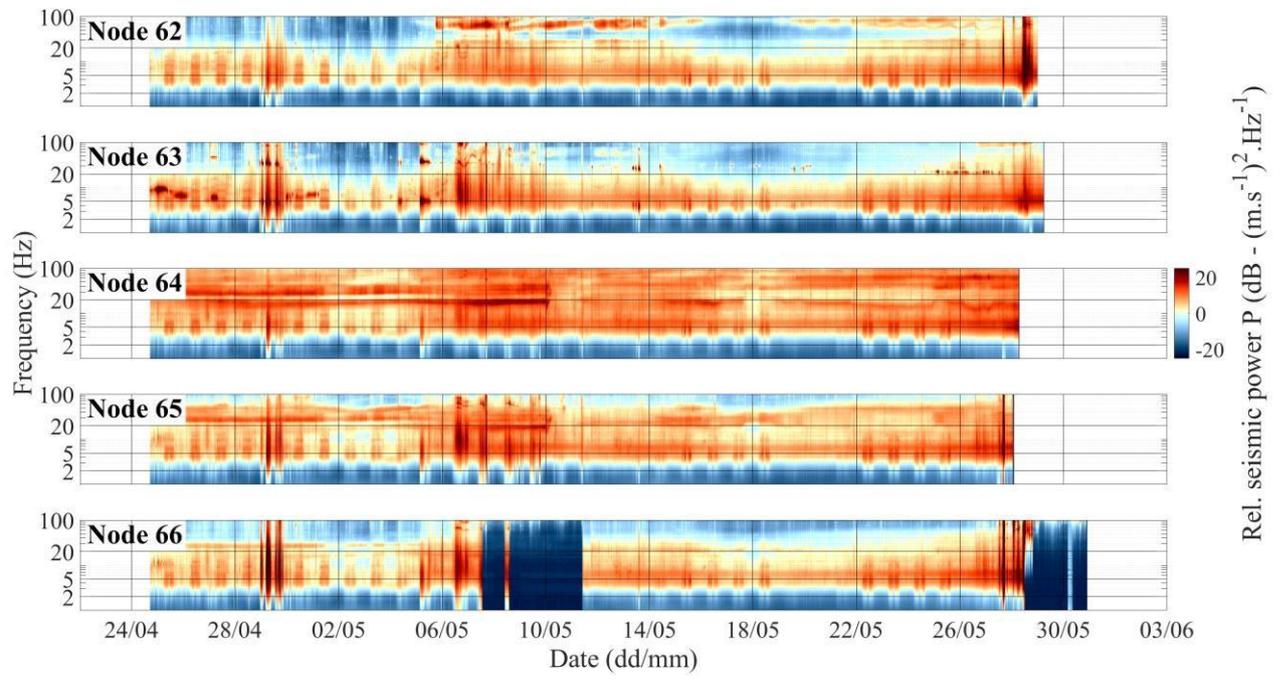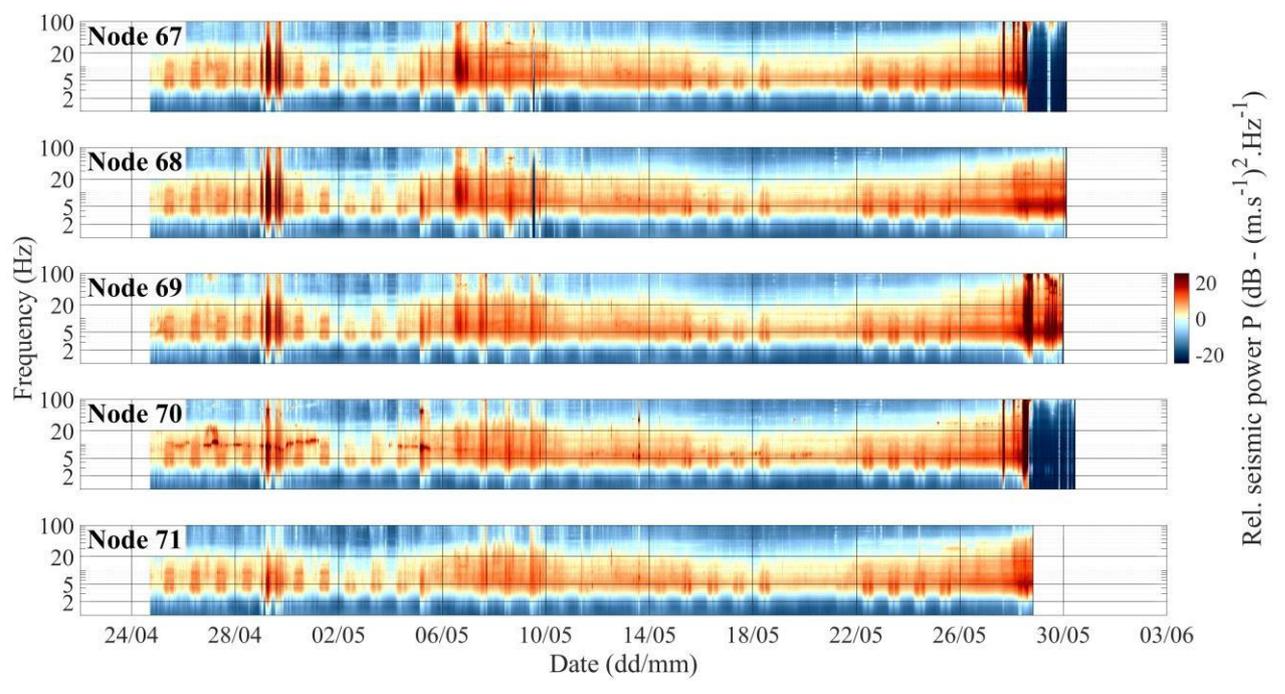

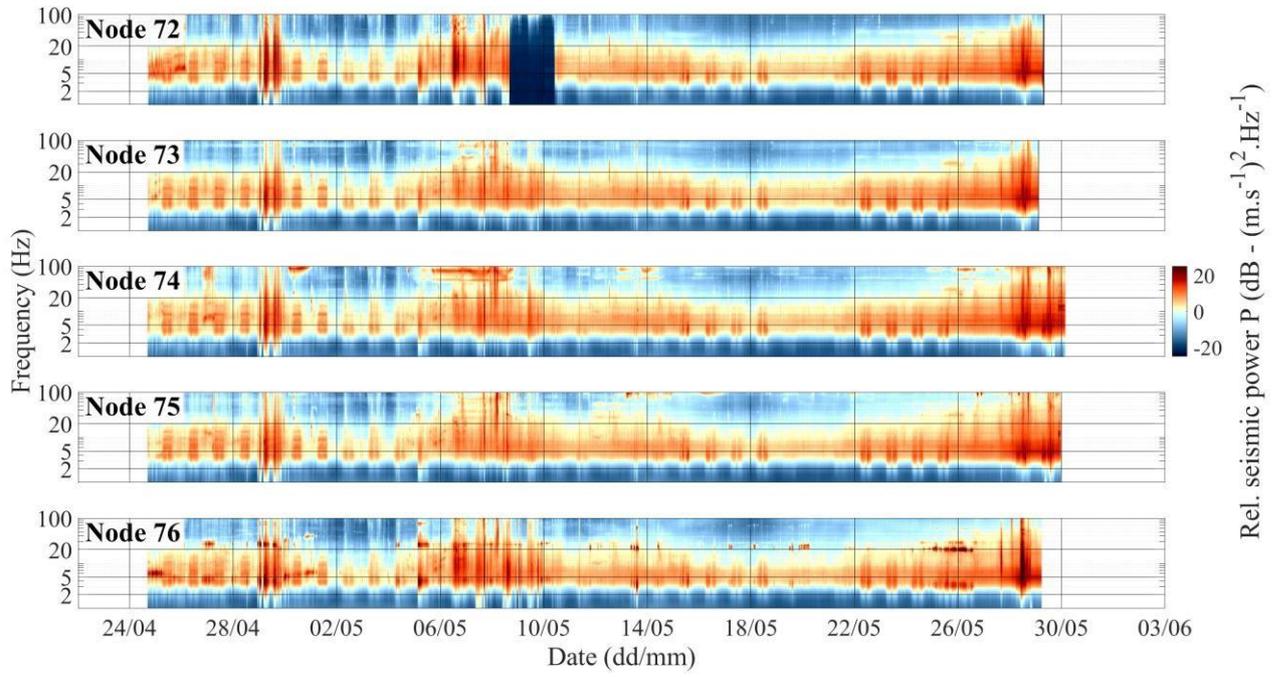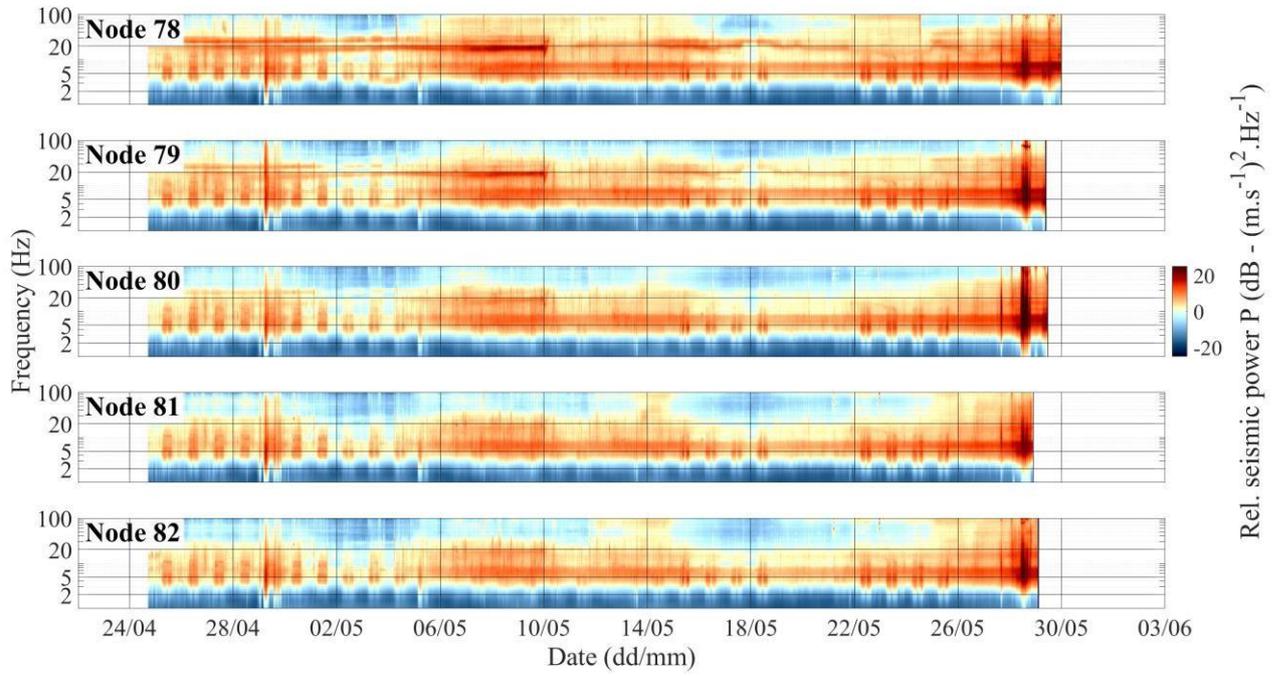

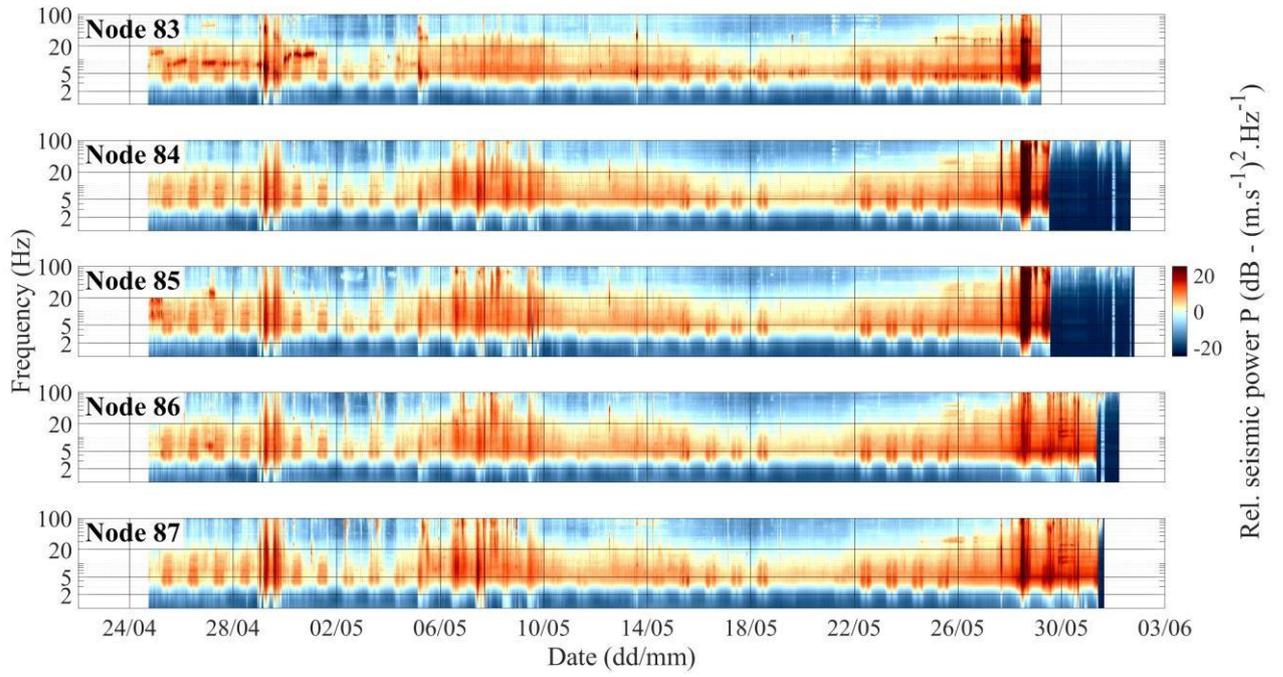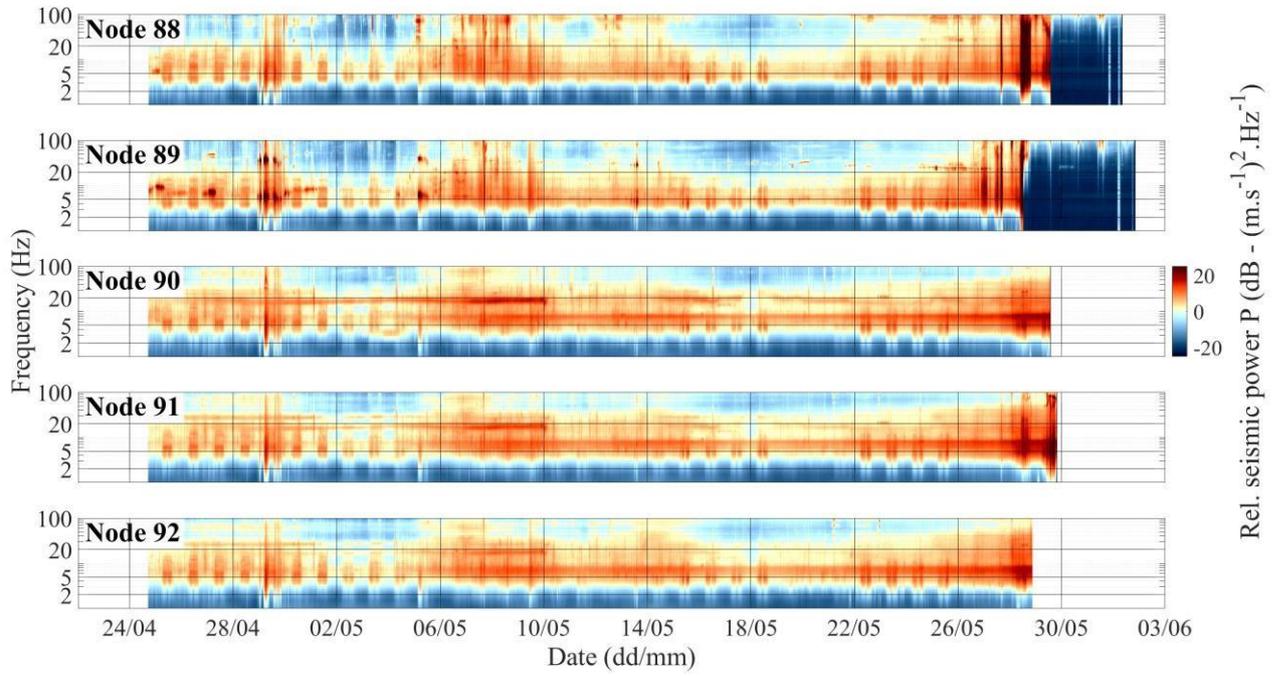

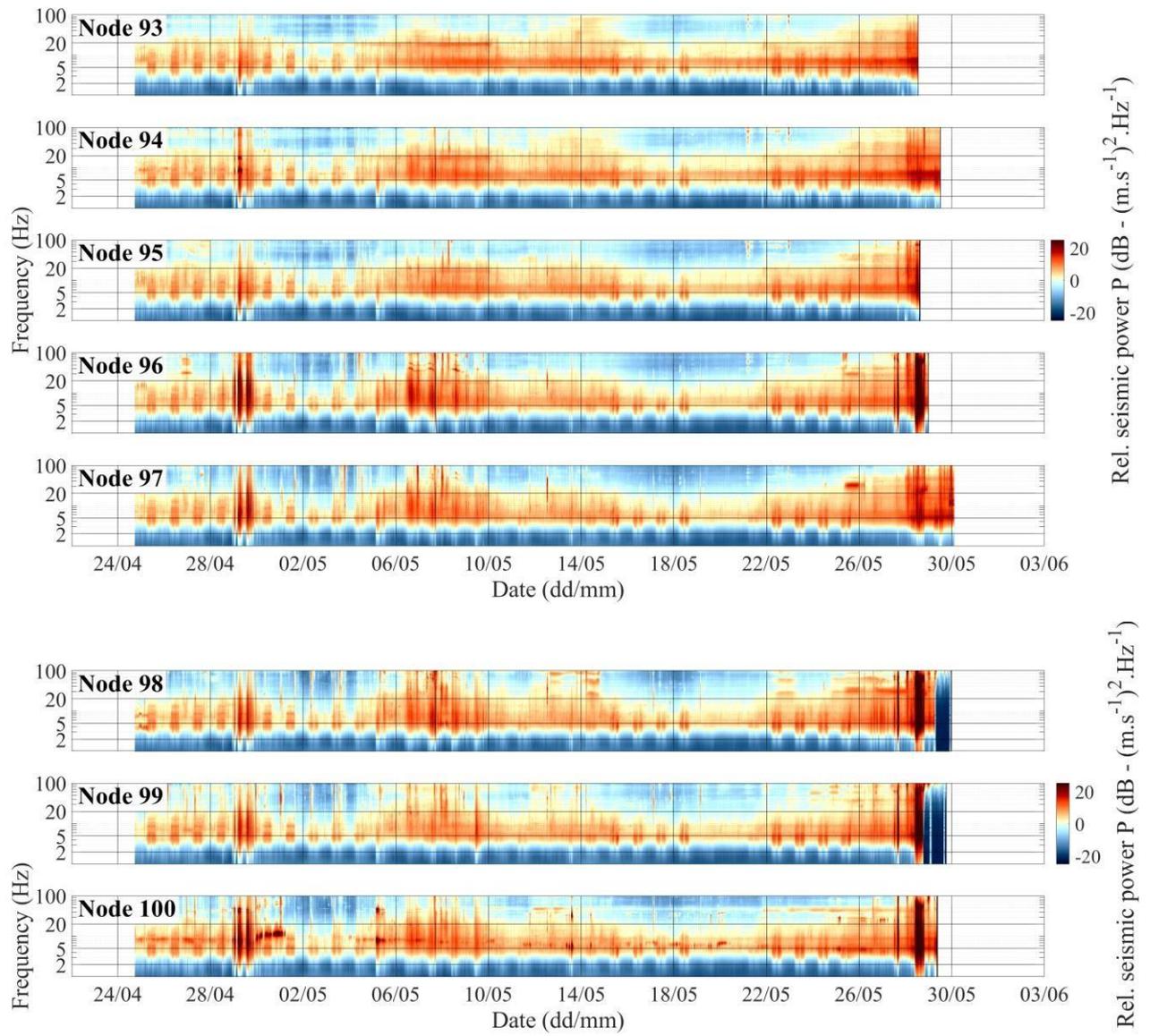

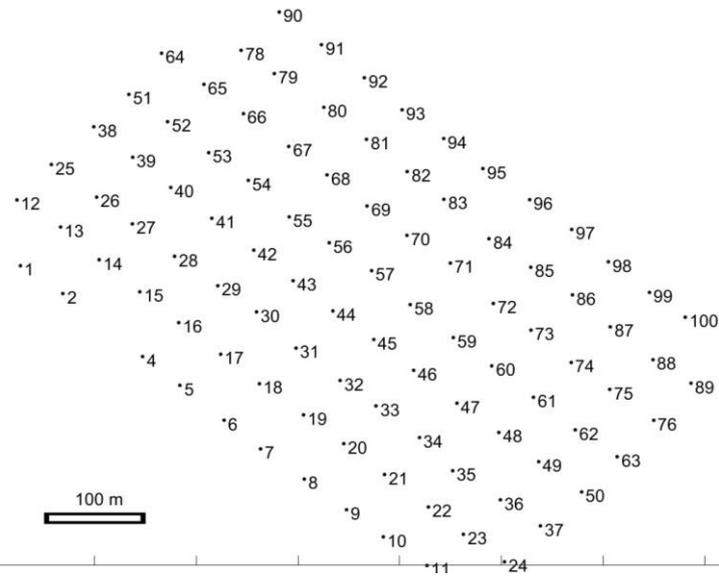

Fig. S1: Spectrograms calculated at all stations over the whole period. Lower panel shows a map view of the nodes nomenclature. The reader should refer to Fig. 1 for absolute positioning of the dense array. Colour scale are equals over all panels and represent the seismic power as calculated in Fig. 4.